\documentclass[a4paper]{article}

\usepackage{ulem}
\usepackage{authblk}
\usepackage{graphicx}
\usepackage{amsmath}
\usepackage{amssymb}
\usepackage{caption}

\usepackage{fullpage}

\begin{document}

\title{Lyapunov analysis of multiscale dynamics:\\ The slow bundle of the two-scale Lorenz '96 model}


\author[1]{Mallory Carlu}{}
\author[1]{Francesco Ginelli}{}
\author[2,3,4]{Valerio Lucarini}{}
\author[1]{Antonio Politi}{}

\affil[1]{SUPA,Institute for Complex Systems and Mathematical Biology, King's College, University of Aberdeen, Aberdeen UK}
\affil[2]{Department of Mathematics and Statistics, University of Reading, Reading, UK}
\affil[3]{Centre for the Mathematics of Planet Earth, University of Reading, Reading, UK}
\affil[4]{CEN, University of Hamburg, Hamburg, Germany}








\maketitle

\begin{abstract}
We investigate the geometrical structure of instabilities in the two-scales Lorenz '96 model through the prism of Lyapunov analysis. Our detailed study of the full spectrum of covariant Lyapunov vectors reveals the presence of a \textit{slow bundle} in tangent space, composed by a set of vectors 
with a significant projection on the slow degrees of freedom; they correspond to the smallest (in absolute value) Lyapunov exponents and thereby
to the longer time scales.
We show that the dimension of the slow bundle is extensive in the number of both slow and fast degrees of freedom, and discuss its 
relationship with the results of a finite-size analysis of instabilities, supporting the conjecture that the slow-variable behavior is effectively
determined by a non-trivial subset of degrees of freedom.
More precisely, we show that the slow bundle corresponds to the Lyapunov spectrum region where fast and slow instability rates overlap, 
``mixing'' their evolution into a set of vectors which simultaneously carry information on both scales.
We suggest these results may pave the way for future applications to ensemble forecasting and data assimilations in weather and climate models.
\end{abstract}



\section{Introduction}

Understanding the dynamics of multiscale systems is one of the great challenges in contemporary science both for the theoretical aspects and the applications in many areas of interests for the society and the private sectors. Such systems are characterized by a dynamics that takes place on diverse spatial and/or temporal scales, with interactions between different scales combined with the presence of nonlinear processes. The existence of a variety of scales makes it hard to approach such systems using direct numerical integrations, since the problem is stiff. Additionally, simplifications based on naïve scale analysis, where only a limited set of scales are deemed important and the others outright ignored, might be misleading or lead to strongly biased results. The nonlinear interaction with scales outside the considered range may, indeed, be important as a result of (possibly slow) upward or downward cascades of energy and information.

A crucial contribution for the understanding of multiscale systems comes from the now classic Mori-Zwanzig theory \cite{Zwanzig1960,Zwanzig1961,Mori1974}, which allows one to construct an effective dynamics specialised for the scale of interest, which are, typically, the slow ones. The enthusiasm one may have for the Mori-Zwanzig formalism is partly counterbalanced by the fact that the effective \textit{coarse-grained} dynamics is written in an implicit form, so that it is of limited direct use. More tractable results can be obtained in the limit of an infinite time scale separation between the slow modes of interest and the very fast degrees of freedom one wants to neglect: in this case, the homogeneisation theory indicates that the effect of the fast degrees of freedom can be written as the sum of a deterministic, drift-like, correction plus a stochastic white-noise forcing \cite{Pavliotis08}.   

The climate  provides an excellent example of multiscale system, with dynamical processes taking place on a very large range of spatial and temporal scales. 
The chaotic, forced, dissipative dynamics and the non-trivial interactions between different scales represent a fundamental challenge in predicting and
understanding weather and climate. A fundamental difficulty in the study of the multiscale nature of the climate system comes from the lack of any spectral gap, namely, a clear and well-defined separation of scales. The climatic variability covers a continuum of frequencies \cite{Peixoto1992,L2014}, so that the powerful techniques based on homogeneization theory cannot be readily applied. 

On the other side, there is a fundamental need to construct efficient and accurate parametrizations for describing the impact of small on larger scales in order to improve our ability to predict weather and provide a better representation of climate dynamics. For some time it has been advocated that such parametrizations should include stochastic terms \cite{Palmer2008}. Such a point of view is becoming more and more popular in weather and climate modelling, even if the construction of parametrizations is mostly based on ad-hoc, empirical methods \cite{Franzke2015,Berner2017}. Weather and climate applications have been instrumental in stimulating the derivation of new general results for the construction of paramerizations of multiscale systems, and for understanding the scale-scale interactions. Recent advances have been obtained using: (i) Mori-Zwanzig and Ruelle response theory \cite{Wouters2012,Wouters2013}; (ii) the generalization of the homogenization theory-based results obtained via the Edgeworth expansion \cite{Wouters2017}; (iii) the use of hidden Markov layers \cite{Chekroun2015a,Chekroun2015b} from a data-driven point of view. An extremely relevant possible advantage of using theory-based methods is the possibility of constructing scale-adaptive parametrizations; see discussion in \cite{VissioLucarini}.


Another angle on multiscale systems deals with the study of the scale-scale interactions, which are key to understand instabilities and dissipative processes and the associated predictability and error dynamics.  Lyapunov exponents \cite{PikovskyPoliti_LE}, which describe the linearized evolution of infinitesimal perturbations, are mathematically well established quantities, and seem to be the most natural choice to start addressing this problem. However, as it is well known, in multiscale systems the maximum (or leading) Lyapunov exponent controls only the 
early-stage dynamics of very small perturbations \cite{Lorenz96}. As time goes on, the amplitude 
of the perturbations of the fastest variables start saturating, while those affecting the slowest 
degrees of freedom grow at a pace mostly controlled by the (typically weaker) instabilities characteristic 
of the slower degrees of freedom. While nonlinear tools, such as finite-size Lyapunov exponents \cite{FSLE}, are able to capture the rate of this multi-scale growth, they lack the mathematical rigour of infinitesimal 
analysis.
In particular, they are unable to convey information on the leading directions of these perturbations, 
when they grow across multiple scales -- an essential problem if one wishes 
to investigate at a deterministic level the non-trivial correlations across structures and perturbations 
acting on different scales. 
It is therefore of primary importance to better understand the multiscale and interactive structure of these instabilities and, in particular, to probe the sensitivity of multiscale systems
to infinitesimal perturbations acting at different spatial and temporal scales and different directions.
To this purpose, infinitesimal Lyapunov analysis allows one to compute not only a full spectrum of Lyapunov Exponents (LEs), 
but also their corresponding {\it tangent space} directions, the so called covariant Lyapunov vectors (CLVs) \cite{Ginelli07}.   
CLVs are associated with LEs (in a relationship that, loosely speaking, resembles the eigenvector-eigenvalue pairing) and provide an intrinsic
decomposition of tangent space that links growth (or decay) rates of (small) perturbations to physically based directions in configuration space. 
In principle, they can be used to associate instability time scales (the inverse of LEs) with well defined real-space perturbations or uncertainties.

While this information is gathered at the linearized level, one may nevertheless conjecture that LEs and CLVs associated with the slowest time scales (i.e. the smallest LEs in absolute value) 
can capture relevant information on the large-scale dynamics and its correlations with the faster degrees of freedom. 
In a sense, one may conjecture that the small LEs and the corresponding CLVs could be used to gain access to a non-trivial effective large scale dynamics. See for instance \cite{Norwood2013} where three coupled Lorenz 63 systems are investigated.
Accordingly, the identification of linear instabilities in full multiscale models, is then expected to have practical implications in terms of control and predictability. 

In the following, we will begin to investigate these ideas studying the tangent-space structure of a simple two-scale atmospheric model, the celebrated Lorenz '96 (L96) model first introduced in \cite{Lorenz96}.

The L96 model provides a simple yet prototypical representation of a two-scale system
where large-scale, synoptic variables are coupled to small-scale,
convective variables. The Lorenz 96 model quickly established
as an important testbed for evaluating new methods of data
assimilation \cite{Trevisan2004, Trevisan2010} and stochastic-parametrization schemes \cite{VissioLucarini, Orrell2003, Wilks2006}.
In the latest decade, it also received considerable attention in the statistical physics community \cite{Abramov2008, Hallerberg2010, Lucarini2011, Gallavotti2014}, while
an earlier study -- limited to the stronger instabilities -- has highlighted  the localization properties of the associated CLVs \cite {Herrera2011}.

Our Lyapunov analysis reveals the existence of a nontrivial {\it slow bundle} in tangent space, formed by a set of CLVs -- associated with the smallest LEs -- the only ones with a non-negligible projection on the slow variables. The number of these CLVs is considerably larger than the number of slow variables, and it is {\it extensive} in the number of slow and fast degrees of freedom.
At the same time, the directions associated with highly expanding and contracting LEs are aligned almost exclusively along the fast, small-scale degrees of freedom. 
Moreover, we show that the LE corresponding to the first CLV of the slow bundle (i.e. the most expanding direction within this subspace) approaches the finite-size Lyapunov exponent in a large-perturbation range, where linearization is not generally expected to apply.

Altogether, it should be made clear that the time-scale separation between the slow bundle and the fast degrees of freedom is large but finite and stays finite when the number of degrees of freedom is let to diverge (i.e. it is not a standard hydrodynamics component). Additionally, the stability is not absolutely weak
in the sense of nearly vanishing Lyapunov exponents.

The  paper is organized as follows. Section 2 introduces both the L96 model and the fundamental tools of Lyapunov analysis used in this paper. Evidence for the existence of a slow bundle is presented in section 3. In Section 4, we investigate how this slow structure arises from the superposition of the instabilities of the slow and fast dynamics respectively. Section 5, on the other hand, is devoted to a comparison with results of finite-size analysis. Finally, in Section 6 we discuss our results, further commenting on their generality and proposing future developments and applications.

\section{The Lorenz '96 model: a simple multiscale system}

\subsection{Model definition and scaling considerations}
\label{modelsec}
The L96 model is a simple example of an extended multi-scale system such as the Earth atmosphere. 
Its dynamics is controlled by synoptic variables, characterized by a slow evolution over large scales, coupled
to the so-called convective variables characterized by a faster dynamics over smaller scales.

The synoptic variables $X_k$, with $k=1,...,K$, represent generic observables on a given latitude circle; each 
$X_k$ is coupled to a subgroup of $J$ convective variables $Y_{k,j}$, ($j=1,...,J$) that follow the faster convective 
dynamics typical of the $k$ sector,
\begin{subequations}
\label{Eq::Original}
\begin{align}
\label{LorenzX}
\dot{X}_k=X_{k-1}(X_{k+1}-X_{k-2}) - X_k + F_s - \frac{hc}{b} \sum_j Y_{k,j} \\
\label{LorenzY}
\dot{Y}_{k,j}=cbY_{k,j+1}(Y_{k,j-1}-Y_{k,j+2})- c Y_{k,j} + \frac{c}{b}F_f + \frac{hc}{b} X_k 
\end{align}
\end{subequations}
In both sets of equations, the nonlinear nearest-neighbor interaction provides an account of advection due to the movement of air masses,
while the last terms describe the mutual coupling between the two sets of variables. Each $X_k$ variable is  affected by the sum of the 
associated $Y_{k,j}$ variables, while each $Y_{k,j}$ is forced by the variable  $X_k$ corresponding to the same 
sector $k$. Finally, the linear terms $-X_k$ and $-cY_{k,j}$ account for internal dissipative processes 
(viscosity) and are responsible for the contractions of the phase space.

We remark that in our configuration, following \cite{VissioLucarini}, energy is injected in the system both at  large and at small scales, provided by the constant terms $F_s$ and $F_f$, which impact the slow and fast scales of the system, respectively.

The presence of the additional forcing term acting on the $Y_{k,j}$ variables makes it possible to have chaotic dynamics on the small scales also in the limit of  vanishing coupling ($h \to 0$), as opposed to the typical L96 setting, where the small-scale variables become spontaenously chaotic without the need of being forced by their associated $X_k$ as a result of downward energy cascade from the slow variables. 

Moreover, the parameter $c$ controls the time-scale separation between the $X_k$ and $Y_{k,j}$ variables, 
while $b$ controls the relative amplitude of the $Y_{k,j}$ components.
 Finally, $h$ gauges the strength of the coupling between slow and fast 
variables.

The L96 model thus contains $K$ slow variables and $K \times J$ fast variables, for a total of $N=K(1+J)$ 
degrees of freedom. It is complemented by the boundary conditions
\begin{gather}
X_{k-K}=X_{k+K}=X_k \nonumber\\
Y_{k-N,j}=Y_{k+K,j}=Y_{k,j}\\
Y_{k,j-J}=Y_{k-1,j}\nonumber\\
Y_{k,j+J}=Y_{k+1,j}\nonumber
\end{gather}
In his original work \cite{Lorenz96}, Lorenz considered $K=36$ slow variables and $J=10$ fast variables for each subsector, for a total of $N=396$ degrees of freedom. As usual, one is ideally interested in dealing with arbitrarily large $K$ and $J$ values,
so that it is preferable to formulate the model in such a way that it remains meaningful in the limit $K,J \to \infty$.
In this respect, the only potential problem is the global coupling, represented by the sum 
in Eq.~(\ref{LorenzX}), which should stay finite for $J\to\infty$. 
This can be easily ensured by imposing that the coefficient in front of the sum is inversely proportional
to $J$.
The most compact representation is obtained by introducing the rescaled variables
$Z_{k,j}=b Y_{k,j}$, and replacing $b$ with a new parameter $f$
\begin{equation}
f=\frac{Jc}{b^2} \,,
\label{Eq::f}
\end{equation}
With these transformations, Eqs.~(\ref{Eq::Original}) can be rewritten as
\begin{subequations}
\label{Eq::Rescaled}
\begin{align}
\label{RescaledX}
\dot{X}_k=X_{k-1}(X_{k+1}-X_{k-2})  - X_k + F_s -  hf \left< Z_{k,j} \right>_j\\
\label{RescaledZ}
\frac{1}{c}\dot{Z}_{k,j}=Z_{k,j+1}(Z_{k,j-1}-Z_{k,j+2}) - Z_{k,j} + F_f + h X_k 
\end{align}
\end{subequations}
where
\begin{equation}
\left< Z_{k,j} \right>_j=\frac{1}{J}\sum_{j=1}^J Z_{k,j} \,, 
\end{equation}
while the boundary conditions are the same as above. %
In practice $f$ gauges the asymmetry of the interaction between slow and fast variables.
From its definition, it is clear that $f$ strongly depends on the scale separation $b$. For the standard
choice of the parameter values (see below), $f=1$, i.e. the average influence of the fast scales on the slow ones 
is the same as the opposite. 
On the other hand, if we increase the value of $b$, $f\to 0$, which corresponds to a master-slave limit, where the fast variables 
do not affect the slow ones but are actually slaved to them. This makes sense because the 
small-scale variables have extremely small amplitude.
The opposite master-slave limit, perhaps more interesting from a climatological point of view, 
corresponds to taking the $h \to 0$ and $f \to \infty$ limits, while keeping the product $hf$ constant.
In this case, the fast variables follow up to first approximation their own autonomous dynamics, but still drive the slow ones through the 
finite coupling term $ hf\left< Z_{k,j} \right>_j$. In this latter limit, we envision the presence of an upscale energy transfer.

Apart from helping to clarify these master-slave limiting cases, such a reformulation of the model also allows also
to better understand that, in order to maintain a fixed amplitude of the coupling term, it is necessary to keep
$f$ constant when $J$ is varied. Selecting a constant value for the time-scale separation $c$,
we choose to rescale $b$ with $J$, as follows:
\begin{equation}
b=\sqrt{\frac{Jc}{f}}\,.
\label{Eq::scaling1}
\end{equation}
With reference to the Lorenz original parameter choices \cite{Lorenz96}, 
$b=c=10$ and $J=10$, we have $f=1$ and the suggested scaling
\begin{equation}
b=\sqrt{10 J}\,.
\label{Eq::scaling2}
\end{equation}
It is finally interesting to note that, in the absence of forcing and dissipation, Eqs.~(\ref{Eq::Original}) reduce to
\begin{subequations}
\label{Eq::Conserved}
\begin{align}
\label{ConsX}
\dot{X}_k=X_{k-1}(X_{k+1}-X_{k-2}) - \frac{hc}{b} \sum_j Y_{k,j} \\
\label{ConsY}
\dot{Y}_{k,j}=cbY_{k,j+1}(Y_{k,j-1}-Y_{k,j+2}) + \frac{hc}{b} X_k 
\end{align}
\end{subequations}
which conserve a quadratic form of slow and fast variables~\cite{VissioLucarini}
\begin{equation}
E= \sum_k X_k^2 + \sum_{k,j} Y_{k,j}^2 = \sum_k \left( X_k^2 + \frac{f}{c} \left< Z_{k,j}^2 \right>_j \right)
\label{energy}
\end{equation}
This conservation law, of course, does not hold in the more interesting forced and dissipative case. However, this result 
suggests that $E$ can be identified with a bona fide energy -- and represents a natural norm -- 
also in the forced and dissipative case.
Note also that, according to the last equality in Eq. (\ref{energy}), changing the number of fast variables does not change the total energy budget, provided that the ratio $f/c$ remains constant. \\

In this study, unless otherwise specified, we will set $f=1$ and typically adopt the slow forcing and the time-scale separation originally adopted by Lorenz, $F_s = 10$ and $c=10$, and choose values for $b$ and $J$ that satisfy the scaling condition (\ref{Eq::scaling2}). According to Lorenz's original derivation, one time unit in this model dynamics is roughly equivalent to 5 days in the real climate evolution \cite{Lorenz96}. \\ 
We will fix $F_f=6$, which guarantees chaoticity in the uncoupled fast variables in the absence of coupling.
Lorenz's original choice for the coupling between the slow and fast scales was $h=1$, but here we will also explore the weak coupling regime, considering coupling values as small as $h=1/16$. 

\subsection{Elements of Lyapunov analysis: Lyapunov Exponents and Covariant Lyapunov Vectors}
As mentioned above, the right tools to quantify rigorously the rate of divergence (or convergence) of nearby trajectories, are the Lyapunov characteristic exponents (LEs) and their associated covariant Lyapunov vectors (CLVs). 
We provide here a qualitative description of these objects. For a more thorough discussion, the reader can look at \cite{Ruelle79, Eckmann85, Ginelli13, Kuptsov12} and references therein.\\
For definiteness, let us consider an $N$ dimensional continuous-time dynamical system 
\begin{equation}
\dot{\mathbf{x}}(t)=\mathbf{f}(\mathbf{x}(t)) \,,
\end{equation}
with $\mathbf{x}(t)$ being the state of the system at time $t$. One can linearize the dynamics around a given trajectory, thus obtaining the evolution of an infinitesimal perturbation $\delta \mathbf{x}(t)$ in the so-called \textit{tangent space}
\begin{equation}
\delta \dot{\mathbf{x}}(t) = \mathbf{J}({\bf x},t) \delta\mathbf{x}(t) \,,
\label{Eq::linear} 
\end{equation}
where we have introduced the {\it Jacobian} matrix
\begin{equation}
\mathbf{J}({\bf x},t) = \frac{\partial {\bf f}({\bf x}(t))}{\partial {\bf x}(t)} \,.
\label{Eq::Jacob} 
\end{equation}
LEs $\lambda_i$ measure the (asymptotic) exponential rates of growth (or decay) of infinitesimal perturbations along a given trajectory. Their plurality holds in the fact that the growth rates associated with different directions of the infinitesimal perturbations are in general different. We then refer to 
the ordered sequence $\lambda_1 \geq \lambda_2 \geq ... \geq \lambda_N$  as to the spectrum of characteristic LEs, or Lyapunov Spectrum (LS), with $N$ being the dimension of the dynamical system. At each point ${\bf x}(t)$ of the attractor, the CLVs ${\bf v}_i({\bf x}(t))$ give the directions of growth of perturbations associated with the corresponding Lyapunov exponent\footnote{In the presence of $m>1$ degenerate (i.e. identical) LEs, the corresponding $m$ CLVs span an $m$-dimensional Oseledets subspace whose elements are all characterized by the same growth rate.}. In other words, they span the Oseledets splitting~\cite{Eckmann85}, i.e.
an infinitesimal perturbation $\delta {\bf x}_i(t_0)$ exactly aligned with the $i$th CLV ${\bf v}_i({\bf x}(t_0))$, after a sufficiently long time $t$ will grow or decay as
\begin{equation}
\left\lVert\delta {\bf x}_i(t_0+t) \right\rVert \approx \left\lVert \delta {\bf x}_i(t_0) \right\rVert e^{\lambda_i t} \,.
\label{Eq::growth} 
\end{equation}
LEs are global quantities, measuring the average exponential growth rate along the attractor, while CLVs are local objects, defined at each point of the attractor and transforming covariantly along each trajectory, according to the linearized dynamics (\ref{Eq::linear}), 
\begin{equation}
\mathbf{v}({\bf x}(t)) = \mathbf{M}({\bf x}_0,t) \mathbf{v}({\bf x}_0) \,,
\label{Eq::linear2} 
\end{equation}
where ${\bf x}_0 \equiv {\bf x}(0)$ and the {\it tangent linear propagator} $\mathbf{M}({\bf x}_0,t)$ satisfies
\begin{equation}
\dot{\mathbf{M}}({\bf x}_0,t)= \mathbf{J}({\bf x},t) \mathbf{M}({\bf x}_0,t) \,,
\label{Eq::M}
\end{equation}
with $\mathbf{M}({\bf x}_0,0)$ being the identity matrix.\\
In the following, we always refer to CLVs assuming they have been properly normalized.
With the above mentioned exception of degeneracies, CLVs constitute an 
intrinsic (they do not depend on the chosen norm) tangent-space decomposition into
the stable and unstable directions associated with the different LEs. 
LEs themselves have units of inverse time, so that the largest positive (in absolute value) exponents -- and their associated CLVs -- describe fast growing (or contracting) perturbations, while the smaller ones corresponds to longer time scales.\\
Unfortunately, Eqs.~(\ref{Eq::growth}-\ref{Eq::linear2}) cannot be used to compute directly any LEs or CLVs beyond the first one. 
Unavoidable numerical errors generated while handling higher-order CLVs are amplified according to a rate dictated by the largest LE, so that
any tangent-space vector quickly converges to the first CLV.
In order to avoid this collapse, it is customary to periodically orthonormalize the vectors with a QR-decomposition~\cite{Shimada79, Benettin80}.
LEs are thereby computed as the logarithms of the basis vectors normalization factors, time-averaged along the entire trajectory.\\
The mutually orthogonal vectors, obtained as a by-product of this procedure, constitute a basis in tangent space and are usually referred to as 
Gram-Schmidt vectors (by the name of the algorithm used to perform the QR-decomposition) or Backward Lyapunov Vectors 
(BLVs, because they are obtained by forward integrating the system until a given point in time, thus spanning the past trajectory with respect to this point). 
Being forced to be mutually orthogonal, BLVs allow only reconstructing the orientation of the subspaces spanned by the most expanding directions.
In this work, we concentrate on the CLVs for the identification of the various expanding/contracting direction.
This is done by implementing a dynamical algorithm, based on a clever combination of both forward and backward iterations of the tangent 
dynamics, introduced in \cite{Ginelli07} and more extensively discussed in Ref.~\cite{Ginelli13}. \\  
In practice, one first evolves the forward dynamics, following a phase space trajectory to compute the full LS $\{\lambda_i\}_{i=1,\ldots,N}$ 
and the basis of BLVs $\{{\bf g}_i(t_m)\}_{i=1,\ldots,N}$ with a series of QR-decompositions performed along the trajectory every $\tau$ time units, 
at times $t_m=m\tau$, with $m=1,\ldots,M$. One is then left with a series of orthogonal matrices ${\bf Q}_m$, whose columns are the BLVs ${\bf g}_i(t_m)$ 
and the upper triangular matrices ${\bf R}_m$ which contain the vector norms and their mutual projections.\\
The key idea is then to project a generic tangent space vector ${\bf u}(t_m)$ on the covariant subspaces $S_j (t_m)$ spanned by the first $j$ BLVs at times 
$t_m$. It can be easily shown \cite{Ginelli13} that this projection, evolved backward in time 
according to the inverse tangent-space dynamics, converges exponentially fast to the $j$th covariant 
vector.\footnote{Or, in the case of degenerate LEs, to a vector belonging to the corresponding Oseledets subspace.}
In practice, this backward procedure can be performed by expressing the CLVs in the BLVs basis, 
\begin{equation}
 {\bf v}_j(t_m) = \sum_{i=1}^j c_{i,j} (t_m) {\bf g}_i(t_m) \,.
\end{equation}
The coefficents $c_{i,j} (t_m)$ thus compose an upper triangular matrix ${\bf C}_m$, whose dynamics is actually determined by the ${\bf R}_m$ 
matrices obtained from the QR-decomposition
\begin{equation}
 {\bf C}_m =  {\bf R}_m  {\bf C}_{m-1}\,.
\end{equation}
This last relationship is easily invertible, assuring a computationally efficient and precise method to follow the backwards dynamics. 

\subsection{Lorenz '96 tangent-space dynamics and algorithmic aspects}

The tangent space dynamics of L96 can be readily obtained by linearizing the phase space evolution equation~(\ref{Eq::Original}),
\begin{subequations}
\begin{align}
\label{linearX}
\delta\dot{X}_k = \delta X_{k-1} \; (X_{k+1}-X_{k-2}) + X_{k-1} \; (\delta X_{k+1}- \delta X_{k-2})-\delta X_k-\frac{hc}{b}\sum_j \delta Y_{k,j} \\
\label{linearY}
\delta\dot{Y}_{k,j} = cb \big[ \delta Y_{k,j+1} \; (Y_{k,j-1}-Y_{k,j+2}) + Y_{k,j+1} \; (\delta Y_{k,j-1}- \delta Y_{k,j+2}) \big]-c \delta Y_{k,j}+\frac{hc}{b}\sum_j \delta X_k \,,
\end{align}
\label{Eq::Tangent}
\end{subequations}
where $\delta X_k$ and $\delta Y_{k,j}$ are infinitesimal perturbations of, respectively, slow and fast variables. Together, they define the tangent space vector 
${\bf u} \equiv (\delta X_1,\ldots, \delta X_K, \delta Y_{1,1}, \ldots \delta Y_{K, J})$. One can easily deduce the Jacobian matrix from Eqs.~(\ref{Eq::Tangent}).\\
In this paper, we numerically integrate Eqs.~(\ref{Eq::Original},\ref{Eq::Tangent}) using a Runge-Kutta 4th order algorithm, with a time step $\Delta t = 10^{-3}$, shorter than the choice $\Delta t = 5 \cdot 10^{-3}$ typically made for 
the standard L96 model. In fact, we have verified that such a small time step is actually required in order to compute the entire spectrum of LEs and CLVs with a sufficient accuracy. Typically, to discount transient effects in numerical simulations, we discard the first $10^3$ time units, split in two equal parts: the first 500 time-units allow for the phase space trajectory to reach its attractor, while the second is used for the convergence of the tangent-space vectors towards the BLVs basis. Afterwards, a forward integration of typically $T=10^3$ time units is performed in order to analyze the properties of tangent space. Due to the highly unstable nature of the L96 model (we will see in the following 
that the maximum LE is around 20, for our choice of parameter values), we have to perform the tangent space orthonormalization every $\tau=10^{-2}$ time units.
Finally, a transient of $10^2$ time units is used during the backward dynamics to ensure the convergence of the backwards vectors to the true CLVs. We have also carefully verified that the forward and backward transients are long enough to guarantee a sufficiently accurate convergence to the true LEs and CLVs.

\subsection{The Lorenz '96 Lyapunov spectrum}
\begin{figure}[t!]
\centering
 \includegraphics[width=0.75\textwidth]{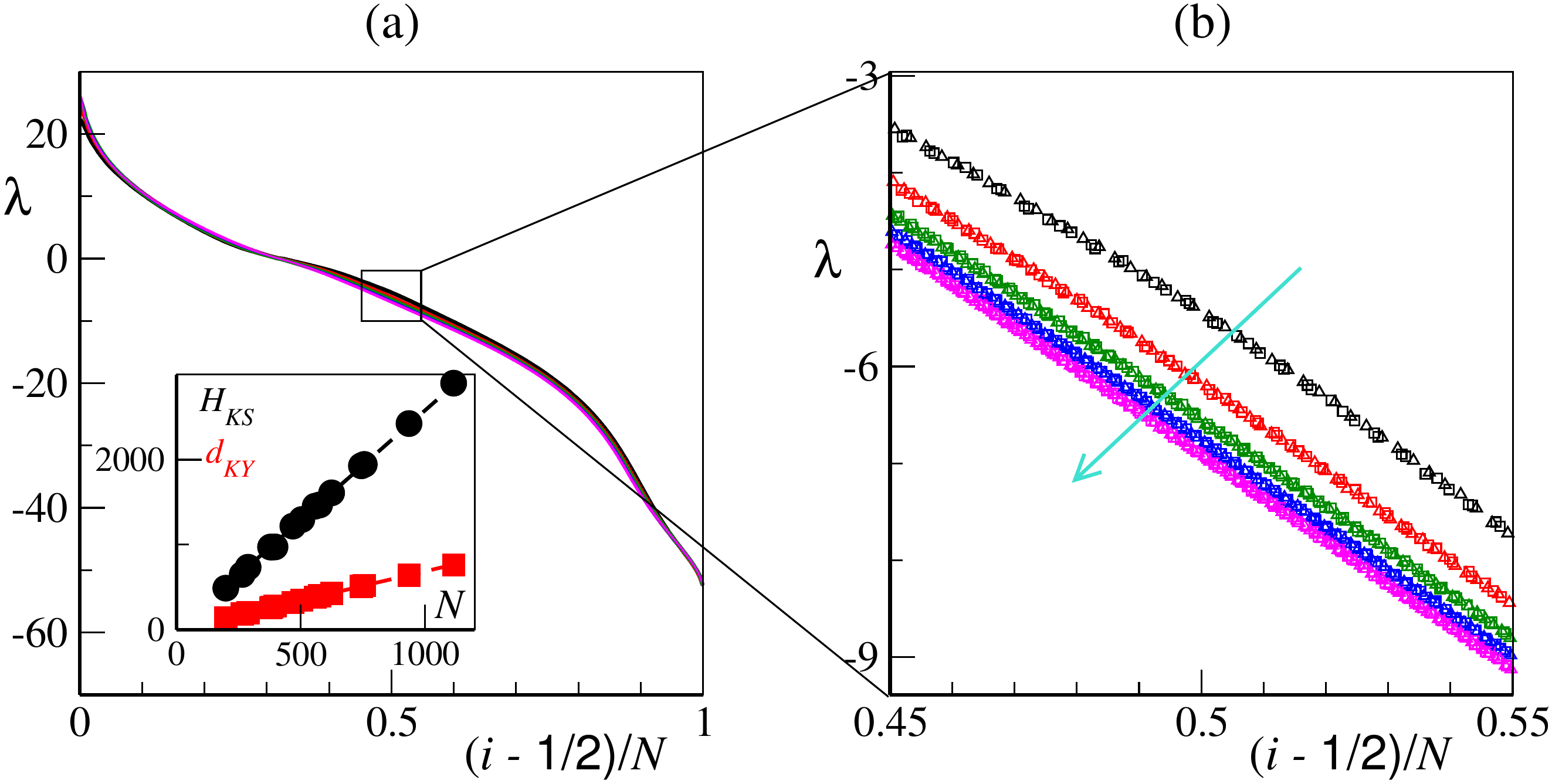}
\vspace{0.5cm}
 \caption{Extensivity of chaos. 
(a) Lyapunov spectra as functions of the rescaled index $i_r=(i-1/2)/N$ for $K=18,24,36$ and $J=10,15,20,25,30$ (all possible combinations). (b) Detail of the central region $i_r \in (0.45, 0.55)$. The cyan arrow marks the direction of increasing $J$ values, while each $J$-branch is the superposition of the spectra for $K=18$ (circles), $K=24$ (squares) and $K=36$ (triangles). Inset of panel (a): Kolmogorov-Sinai entropy $H_{KS}$ (black circles) and  Kaplan-Yorke dimension $d_{KY}$ (red squares) as a function of the number of degrees of freedom $N=K(J+1)$. 
The dashed lines mark a linear fit with zero intercept and slope $\approx 2.6$ ($H_{KS}$) and $\approx 0.7$ ($d_{KY}$).
Simulations have been performed with $h=1$ and $b=\sqrt{10 J}$ (see main text).}
\label{Fig1}
\end{figure}
Spatially extended systems are known to typically exhibit an extensive Lyapunov spectrum \cite{Ruelle78, Livi86, Grassberger89}. 
This property is instrumental for the identification of intensive and extensive observables in the thermodynamic sense. 

Extensivity means that for $N$ tending to infinity (i.e. in the so-called thermodynamic limit), 
the spectrum $\lambda_i$ is a function of the rescaled index 
$\rho = i/N$ only~\footnote{Actually, it is customary to define $\rho$ as $(i-1/2)/N$ to reduce the amplitude of
finite-size corrections~\cite{PikovskyPoliti_LE}.}.

The single-scale L96 model (i.e. Eq.~(\ref{LorenzX})) without the coupling to the fast scale) is no exception \cite{Karimi10, Gallavotti2014}.
Here we show that extensivity of chaotic behavior holds also in the two-scale model provided that -- as discussed in Sec.~\ref{modelsec} -- 
the relation (\ref{Eq::scaling1}) is satisfied.
In the present context, the total number $N$ of degrees of freedom is controlled by two separate indicators, $K$ and $J$, so that,
in principle one can define two distinct thermodynamic limits, i.e. $K\to\infty$ and $J\to\infty$. 
However, in practice, as long as $K,J\gg 1$ the spectral shape depends only on $N$ alone, as seen in
Fig.~\ref{Fig1}, where several different LS nicely overlap. 

In order to appreciate the different role of $K$ and $J$, it is necessary to zoom in, as shown in Fig.~\ref{Fig1}b, where the region
characterized by a larger spread is displayed.
The single spectra are grouped into five different branches, each corresponding to three different values of $K$ 
($K=18, 24, 36$, respectively marked as circles, squares and triangles) and to the same $J$. 
As $J$ increases from 10 to 30, these branches converge to a limiting spectrum, which corresponds to the double thermodynamic
limit $K,J\to\infty$.
The indistinguishability of the spectra obtained for the same $J$ shows that $K$-type finite-size corrections are very small
for given $K$: this is not a surprise, since the number of fast variables, $KJ$ is much larger than that of the slow variables, $K$.

The existence of a limit spectrum implies that the Kolmogorov-Sinai entropy $H_{KS}$ -- a measure of the diversity of the 
trajectories generated by the dynamical system -- is proportional to the number $N$ of degrees of freedom. This can be
appreciated in the inset of Fig.~\ref{Fig1}a (see the black circles), where $H_{KS}$ is determined through the
Pesin formula, which provides an upper bound to $H_{KS}$~\cite{Eckmann85}
\begin{equation}
H_{KS} = \sum_{\lambda>0} \lambda_i  \, .
\label{eq:pesin}
\end{equation}
Similarly, the dimension of the attractor, i.e. the number of ``active'' degrees of freedom, is proportional to $N$, 
as seen again from Fig.~\ref{Fig1}a, where we have plotted the Kaplan-Yorke dimension $D_{KY}$~\cite{Eckmann85} (see red squares)
\begin{equation}
D_{KY} = M + \frac{\sum_{i\le M} \lambda_i}{|\lambda_{M+1}|}  \, ,
\label{eq:KY}
\end{equation}
with $M$ being the largest integer such that $\sum_i \lambda_i >0$.

\begin{figure}[t!]
\centering
 \includegraphics[height=0.30\textheight]{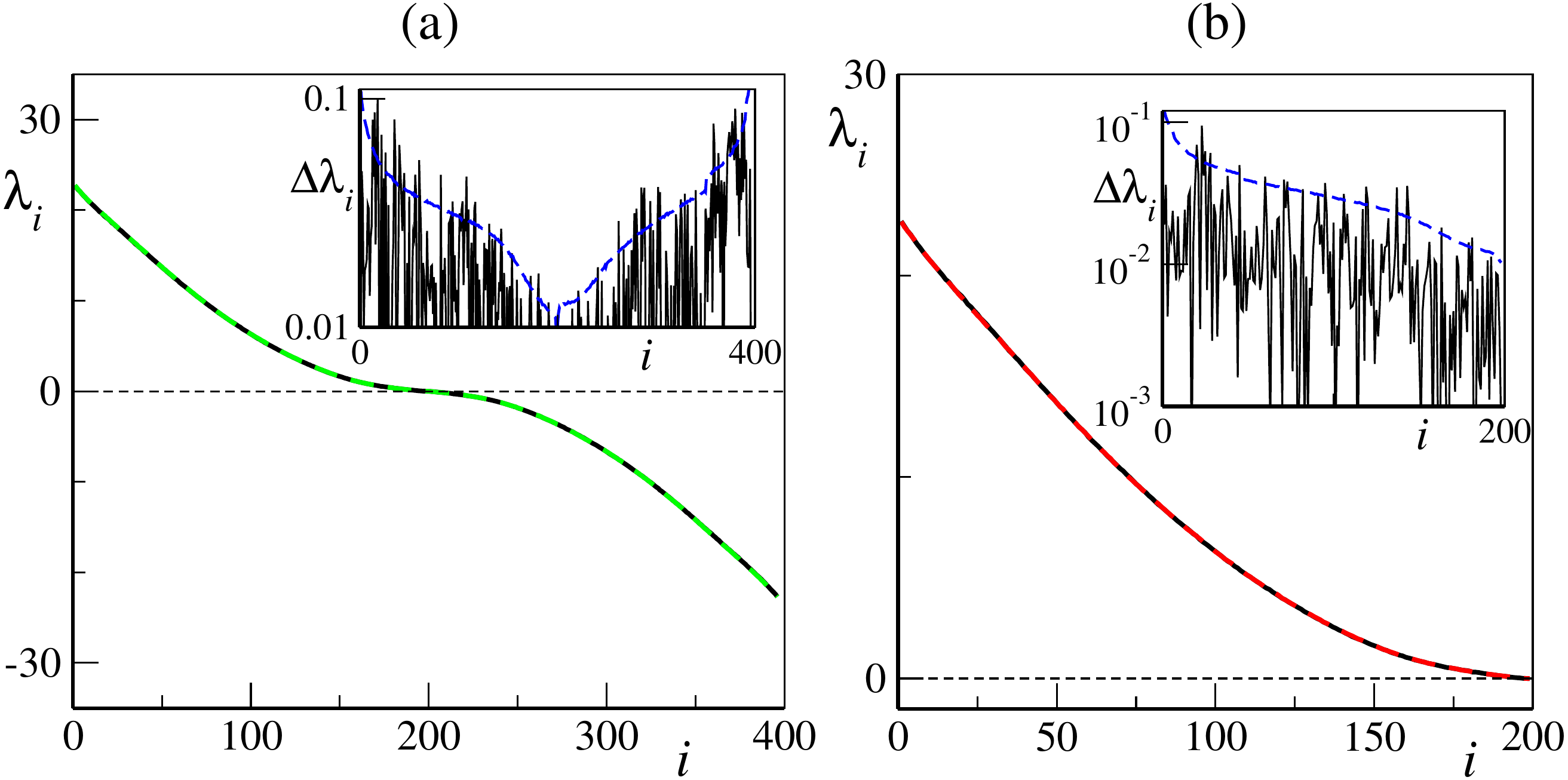}
\vspace{0.1cm}
	\caption{(a) Lyapunov Spectrum for the conservative setup, Eq.~(\ref{Eq::Conserved}). Parameters are $b=10$, $K=36$, $J=10$ ($N=396$ LEs), and initial conditions are chosen such that the conserved energy is $E=36$. The full black line corresponds to $h=1$, while the dashed green line corresponds to $h=-1$. (Inset): The absolute difference $\Delta \lambda_i $ between the two LS is compared with one standard error in our numerical estimate (dashed blue line).
 (b) The second half of the Lyapunov spectrum ($i \in [199, 396]$, dashed red line) is folded under the reflection transformation $\lambda_i \to -\lambda_{N-i+1}$ over its first half ($i \in [1, 198]$, full black line).
(Inset): The absolute difference $\Delta \lambda_i$ between LS and its folded transformation is compared with one standard error in our numerical estimate (dashed blue line).}
\label{Fig2}
\end{figure}

We conclude this section with a brief remark on the Lyapunov spectrum in the zero-dissipation limit 
(\ref{Eq::Conserved}), shown in Fig.~\ref{Fig2}a for $h=1$.
The LS depends only on the initial value of the energy of the system
(in our simulations we have chosen $E=36$ -- see  Eq.~(\ref{energy}))\footnote{The invariant measure is absolutely continuous with respect to the Lebesgue one in the energy shell}.
Moreover, from Fig.~\ref{Fig2}b, we can appreciate that the LS is perfectly symmetric, 
since the second half of the spectrum  superposes to the first half under the transformation 
$\lambda_i \to -\lambda_{N-i+1}$ within numerical precision.  This symmetry is an unexpected general property, 
which holds for any choice of $h$, $c$ and $b$.
In fact, the conservation law can only account for the existence of an extra zero-Lyapunov exponent.
The overall symmetry of the LS must follow from more general properties 
such as the symplectic structure of the model, or
invariance under time-reversal of the evolution equations.
Unfortunately, this model is known to possess no symplectic structure, even if the energy is conserved \cite{Blender2013} and
the only symmetry we have been able to find is the invariance under the transformation
$t \to -t$, $X_k \to -X_k$, $Y_{k,j} \to -Y_{k,j}$, accompanied by a change of the coupling constant $h$.
Indeed, the dashed green line in Fig.~\ref{Fig2}a shows that the Lyapunov spectrum is invariant under the 
transformation $h \to -h$. Therefore, the overall symmetry remains an unexplained property.

\section{Slow tangent-space bundle}

\subsection{Projection of CLVs in the X subspace}
\label{proj}

We now come to the central result of this paper, namely the existence of a nontrivial subspace in tangent space associated with the slow dynamics of the L96 model.

The individual LEs $\lambda_i$ represent the average growth rate (and thus the inverse of suitable time 
scales) of well defined small perturbations aligned along the corresponding CLV, ${\bf v}_i$.
It is therefore logical to ask which of these ``fundamental'' perturbations are more relevant for the
evolution of the accessible macroscopic observables. In the present case, it is natural to focus our
attention on the alignment along the slow variables $X_k$.

The norm of the (rescaled) $i$th CLV can be written as 
\begin{equation}
||\delta X^{(i)}||^2 + ||\delta Y^{(i)}||^2 = 1 \, ,
\label{Eq::normX}
\end{equation}
where the two addenda represent the squared Euclidean norm of the projection onto the slow and fast variables, $\delta X^{(i)} =  (\delta X^{(i)}_1,\ldots, \delta X^{(i)}_K)$
and  $\delta Y^{(i)} =  ( \delta Y^{(i)}_{1,1}, \ldots \delta Y^{(i)}_{K, J})$ respectively.
The most natural indicator of how much the $i$th CLV projects on the slow modes is thus 
the $X$-projected norm $\phi_i \equiv ||\delta X^{(i)}||^2$.

However, it should be noted that, although the CLVs are intrinsic vectors, their orientation
does depend on the relative scales used to represent the single variables and, in particular, fast and slow ones.
If we change the units of measure used to quantify the fast $Y$ variables, introducing
$V_j = \gamma Y_j$, the (Euclidean) norm of the $i$th CLV becomes 
\[
L = \phi_i + \gamma^2(1-\phi_i) \, .
\]
As a result, in the new representation, the weight of the projection on the slow variables becomes
\[
\phi'_i = \frac{\phi_i}{L} \; ,
\]
which shows how the amplitude of the projection depends on the relative scale used to measure 
fast and slow variables.
Since in the very definition of energy (see Eq.~(\ref{energy})), $X$ and $Y$ variables are weighted in
the same way, it is natural to maintain the original definition, i.e. to assume $\gamma=1$.
Nevertheless, as we will see while discussing the evolution of finite perturbations, 
the relative scale is an important parameter we can play with to extract useful information.\\

Given the strong temporal fluctuations of $\phi_i(t)$ when the vectors are covariantly transformed along a 
trajectory (see the end of this section), it is convenient to
refer to its time average (which, assuming ergodicity, corresponds to an ensemble average over the 
invariant measure), 
\begin{equation}
\Phi_i = \langle \phi_i(t) \rangle_t \,.
\label{Eq::normX2}
\end{equation}

\begin{figure}[!t]
\centering
 \includegraphics[width=0.7\linewidth]{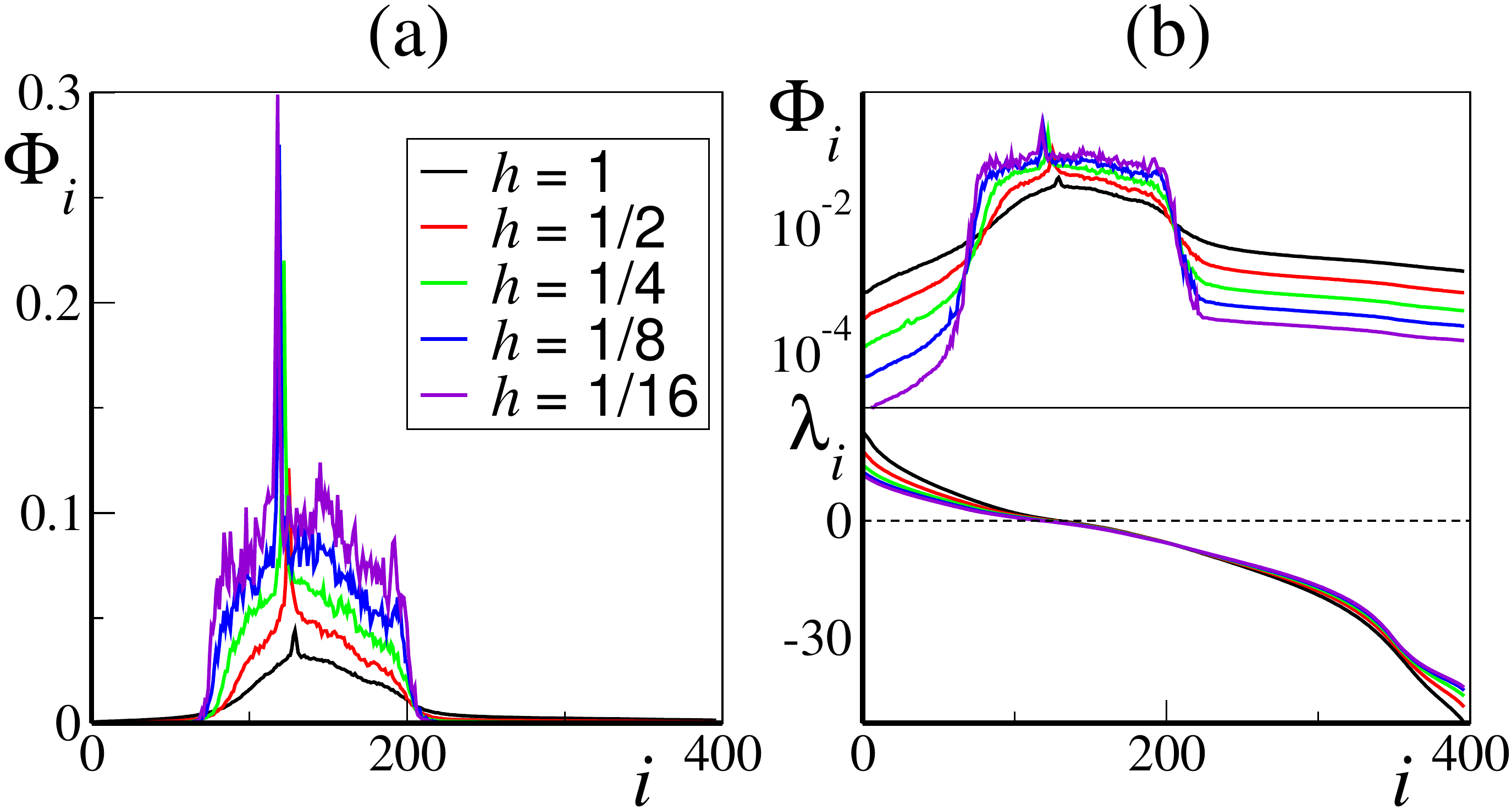}
 \caption{CLVs projection on the slow variables. (a) CLVs average projection norm $\Phi_i$ of the CLVs (see text) as a function of the vector index
for $K=36$, $J=10$ and $b=10$ upon varying the coupling constant $h$.
(b) Upper panel: same as in (a) but in a logarithmic scale vertical scale. Bottom panel: Corresponding Lyapunov Spectra.}
\label{Fig3}
\end{figure}

We have first computed the projection norm $\Phi_i $ for the entire spectrum of vectors in a system of size $K=36$ and $J=10$ with the ``standard'' parameter value $b=10$. In a wide range of coupling strengths -- from strong ($h=1$) to weak ($h=1/16$) -- we find that
both rapidly growing and rapidly contracting perturbations are almost orthogonal to the slow-variable subspace,
the associated CLVs exhibiting a negligible projections over the $X$ directions (see Fig.~\ref{Fig3}a).
In fact, only a ``central band'' constituted by the CLVs associated with the smallest LEs 
display a significative projection over the slow variables. Note however that the typical LEs associated to the ``central band'' CLVs are clearly finite, and are deemed ``small'' only in a relative sense, i.e. when compared with the largest positive and negative exponents of the full spectrum. For instance, for $h=1/4$, we can approximately estimate the corresponding portion of the LS to extend between magnitude $2$ and $-5$. We will comment further on this point in Sec. \ref{sec4}.

Note also that the CLV associated with the only null LE (in the following we simply denote it as the 0-CLV) displays a sharp peak of the projection norm $\Phi_i$. This is just a consequence of the delocalization of this CLV: the perturbation corresponding to the zero exponent points exactly along the trajectory. Direct integration of the phase space equations (not shown) confirms that the total variability of the slow variables is of the same order of magnitude as the total variability of the fast ones.
This central band of CLVs defines the tangent-space {\it slow bundle} relevant for this paper. It becomes more sharply defined for small values of the coupling $h$ but it keeps approximately the same position and width as the coupling $h$ is increased. In particular, for this set of parameter values, 
this non-trivial {\it slow bundle} extends in tangent space over roughly 120 CLVs, much more than the 
$K=36$ slow degrees of freedom.
The extension of the slow bundle can be better appreciated in Fig.~\ref{Fig3}b, where the time-averaged projections are shown in logarithmic scale (top panel) and compared with the full spectrum of LEs (bottom panel).
We are interested in the dependence of this bundle on the number of slow and fast variables. 
As discussed in the previous section, the L96 model is extensive in both the slow and fast variables, provided that the ratio $f= J c /b^2$ is kept 
constant (for the ``standard'' choice of parameters $c=10$ and $f=1$, so that it is sufficient to set $b=\sqrt{10 \, J}$).
In the following we present results for $h=0.5$, but we have carefully verified 
that analogous results hold for other values of the coupling constant $h$.

\begin{figure}[!t]
\centering
 \includegraphics[height=0.55\textheight]{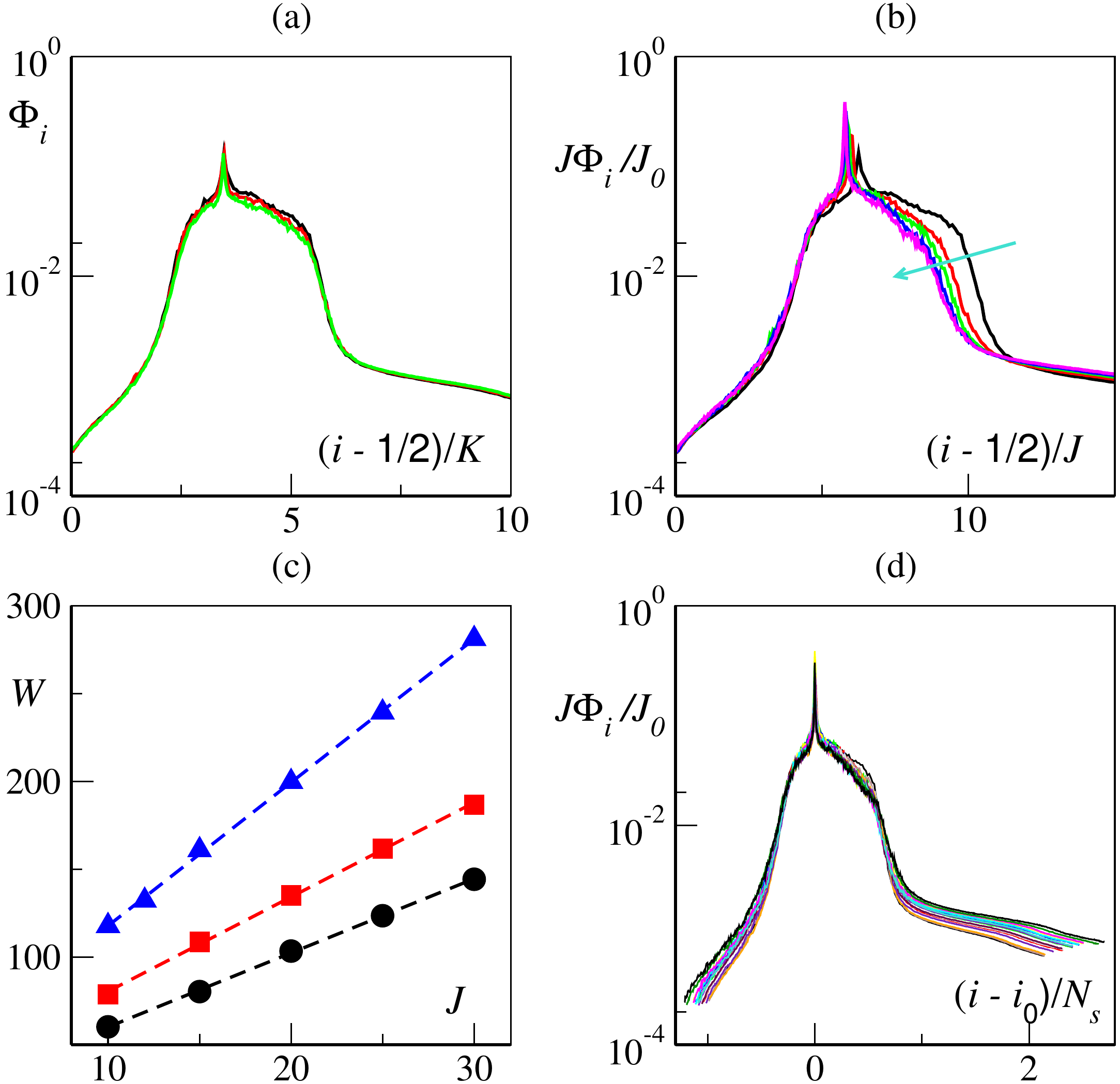}
\vspace{0.1cm}
\caption{Slow bundle scaling for $h=0.5$ and $f=1$. (a) CLVs time-averaged X-projection norm $\Phi_i$ for $K=18,24,36$ and $J=10$ vs. the rescaled index $(i-0.5)/K$. (b) Rescaled (see main text) CLVs average projection norm $\Phi_i$ for $K=18$ and $J=10,15,20,25,30$ (increasing along the cyan arrow) vs. the rescaled index $(i-0.5)/J$. A lack of precise collapse can be appreciated on the rightmost side of the ''central band''.
(c) Central band width $W$ (see main text for details) as a function of $J$ for $K=18$ (black circles), $K=24$ (red squares) and $K=36$ (blue triangles). The best linear fits, marked by the dashed lines are $W=18(1) + 4.20(5) J$ (for $K=18$), $W=26(2) + 5.4(1) J$ ($K=24$) and $W=36(2) + 8.2(2) J$ ($K=36$).  
(d) Rescaled CLVs average projection norm $\Phi_i$ for $K=18,24,36$ and $J=10,15,20,25,30$ (all combinations) vs. the rescaled index $(i-i_0)/N_s$
	Here $i_0$ is the index of the 0-CLV and $N_s = K (1+\alpha J)$, with $\alpha \approx 0.22$ (see text). 
We choose the vertical axis rescaling reference as $J_0=10$.
}
\label{Fig4}
\end{figure}

We first set $J=10$ and explore the behavior of the slow bundle when $K$ is varied
(note that no parameter rescaling is required while changing $K$).
Our simulations, reported in Fig.~\ref{Fig4}a, clearly show that the slow bundle is extensive w.r.t. $K$: upon rescaling the vector index as $i \to (i-0.5)/K$ we observe a clear collapse of the projection patterns.\\
We next focus on the scaling with $J$ at fixed $K$. For the sake of computational simplicity, 
we first consider $K = 18$.
As for the horizontal variable, it is natural to rescale the CLVs index by the number $J$ of fast 
degrees of freedom per subgroup.
Moreover, since $J$ corresponds to the ratio between the number of fast ($K\,J$) and slow ($K$) 
variables and we use the standard Euclidean norm to quantify the projection on the $X$ subspace, 
one expects the fraction $\Phi_i$ of $X$-norm to be inversely proportional to $J$. 
The projection data reported in Fig.~\ref{Fig4}b indeed show a convincing vertical collapse 
of the rescaled $X$-norm $\Phi_i J/J_0$ (here we fix a reference $J_0=10$), but accompanied by a shrinking of the central band on the right side, as $J$ is increased. \\

In order to accurately determine the width of the central band, i.e. the slow bundle dimension, 
we fix a threshold for the {\it rescaled} $X$-norm, $J \Phi_i/J_0= 10^{-2}$, and estimate the number $N_s$ of CLVs with a projection above such a threshold (we have verified that our results hold within a reasonable range of thresholds). The resulting widths $N_s(K, J)$, computed for different numbers of slow variables $K$, are summarized in Fig.~\ref{Fig4}c, where they are plotted versus $J$. For fixed $K$, we see a clear linear increase, 
compatible with the law 
\begin{equation}
N_s(K,J) = K \left(1+\alpha J \right)
\label{Eq::scaling}
\end{equation}
where the coefficient $\alpha$ depends on  the values of $h$, $c$, $F_s$ and $F_f$. In the present case, a
best fit gives $\alpha \approx 0.22$. 
The most general representation of the projections is finally obtained  by 
rescaling the index according to $N_s$ and by using the 0-CLVs (which corresponds to the peak of $\Phi_i$) as
the origin of the horizontal axis.  The excellent collapse in Fig.~\ref{Fig4}d
confirms the extensivity of the slow bundle with both $K$ and $J$. The slow bundle 
is not a simple representation of the $X$ subspace: it does not concide with the slow variables themselves, 
but involves also a finite fraction $\alpha$ of the fast ones, 
singling out a fundamental set of tangent space perturbations closely associated with the slow dynamics.
The origin of the phenomenological scaling law (\ref{Eq::scaling}) will be discussed in the next section.\\

Before concluding this section, we would like to briefly discuss the time-resolved projected norm $\phi_i (t)$. So far, we have discussed time-averaged quantities, but it is worth mentioning that the $X$ projections of individual CLVs are extremely intermittent, hinting at a complex tangent-space flow structure.

\begin{figure}[!t] 
  \centering
  \includegraphics[width=0.8\linewidth]{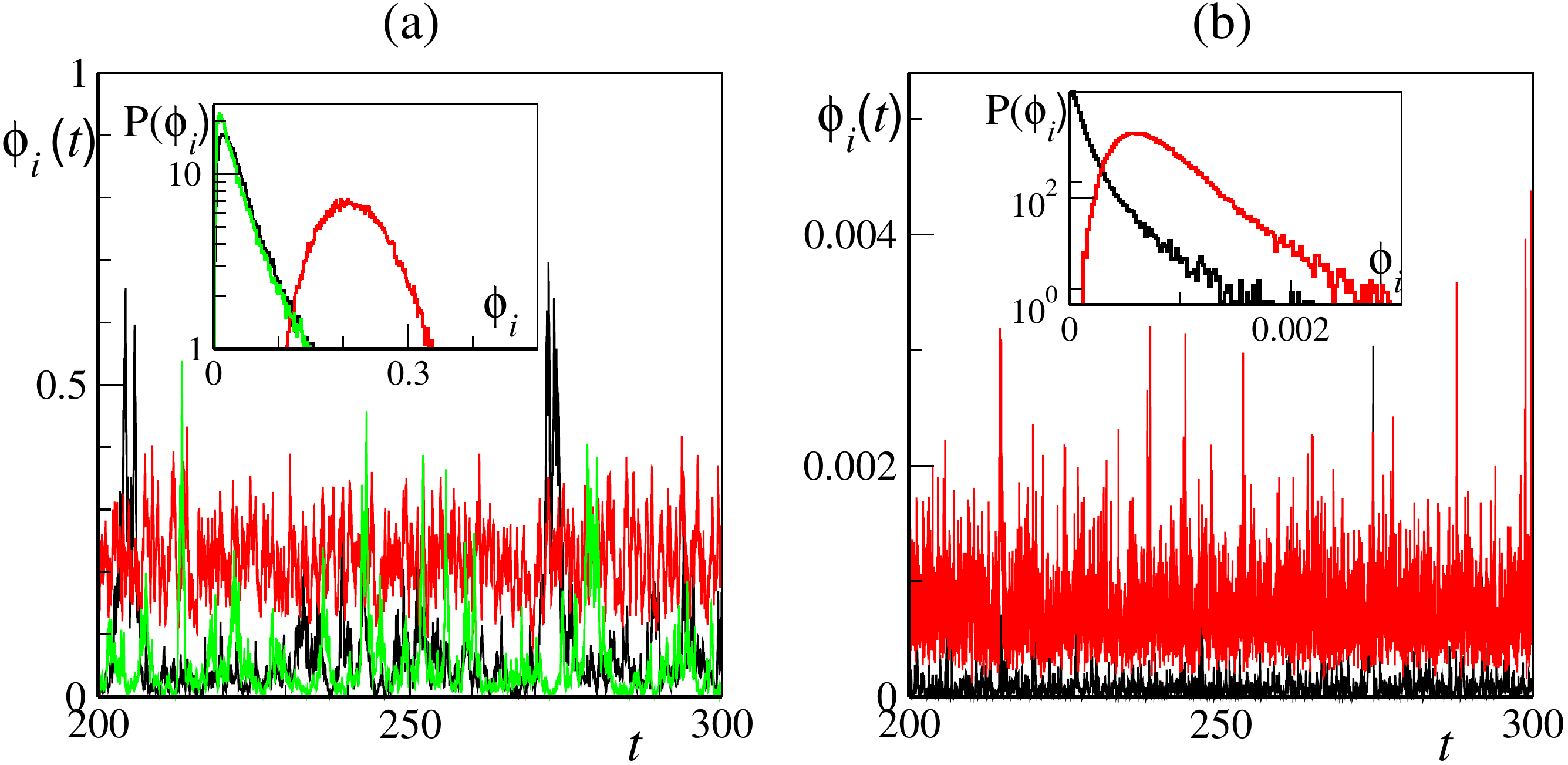} 
\vspace{0.1cm} 
\caption{Time trace and probability distribution of CLV instantaneous projection in the $X$ subspace for $h=1/4$,  $b=10$, and $K=36$, $J=10$. (a) Slow bundle vectors, $i=110$ (black line), $i=122$ (red line, the 0-CLV) and $i=160$ (green line). In the inset: Corresponding probability distributions of the three CLVs time traces.
(b) For two other CLVs with negligible $X$-projection, first vector (black line) and 255th vector(red line). In the inset: Corresponding probability distributions of the two CLVs time traces.}  
  \label{fig::RestricLE_XY}
\end{figure}

In Fig.~\ref{fig::RestricLE_XY}a we display a few selected time series of the norm $\phi_i(t)$ for $b=10$, $K=36$, $J=10$ and $h=1/4$ (the overall picture does not change qualitatively
for different choices of the coupling strength). The time series correspond to:
the 110th vector, located in the left part of the central band (before the 0-CLV),
the 160th vector located in the right part, and the 0-CLV (vector index $i=122$). We clearly see a strong intermittency, resulting in a very skewed distribution of the time resolved $X$ norms. The 0-CLV is an exception, displaying more regular oscillations and a rather symmetric distribution around its mean value. This confirms the peculiar nature of the 0-CLV, whose delocalized structure is essentially determined by its alignment with the phase space flow.

In Fig.~\ref{fig::RestricLE_XY}b we display vectors outside the slow bundle: the 1st and the 250th, which are on the left and on the right hand side of the central band, respectively. We see that also vectors outside the slow bundle display a certain degree of intermittency, albeit on a faster time scale, and rather skewed distributions of their $\phi_i(t)$ values. Their $X$-projection, of course, is strongly suppressed and remains very close to 0. In Sec. \ref{sec4}, we will further comment on the intermittent behavior of $\phi_i(t)$, showing that it arises from near degeneracies in the instantaneous instability rates.

\section{The origin of the slow bundle}
\label{sec4}

In the previous section we have identified a slow bundle in the tangent space of the
L96 model -- a ``central band'' centered around the 0-CLV -- whose covariant vectors 
are characterized by a large projection over the slow degrees of freedom.
It is natural to expect this band to be associated not only with long time scales (i.e. the inverse of the 
corresponding LEs), but also with large scale-instabilities.

\begin{figure}[!h]
\centering
 \includegraphics[width=0.7\textwidth]{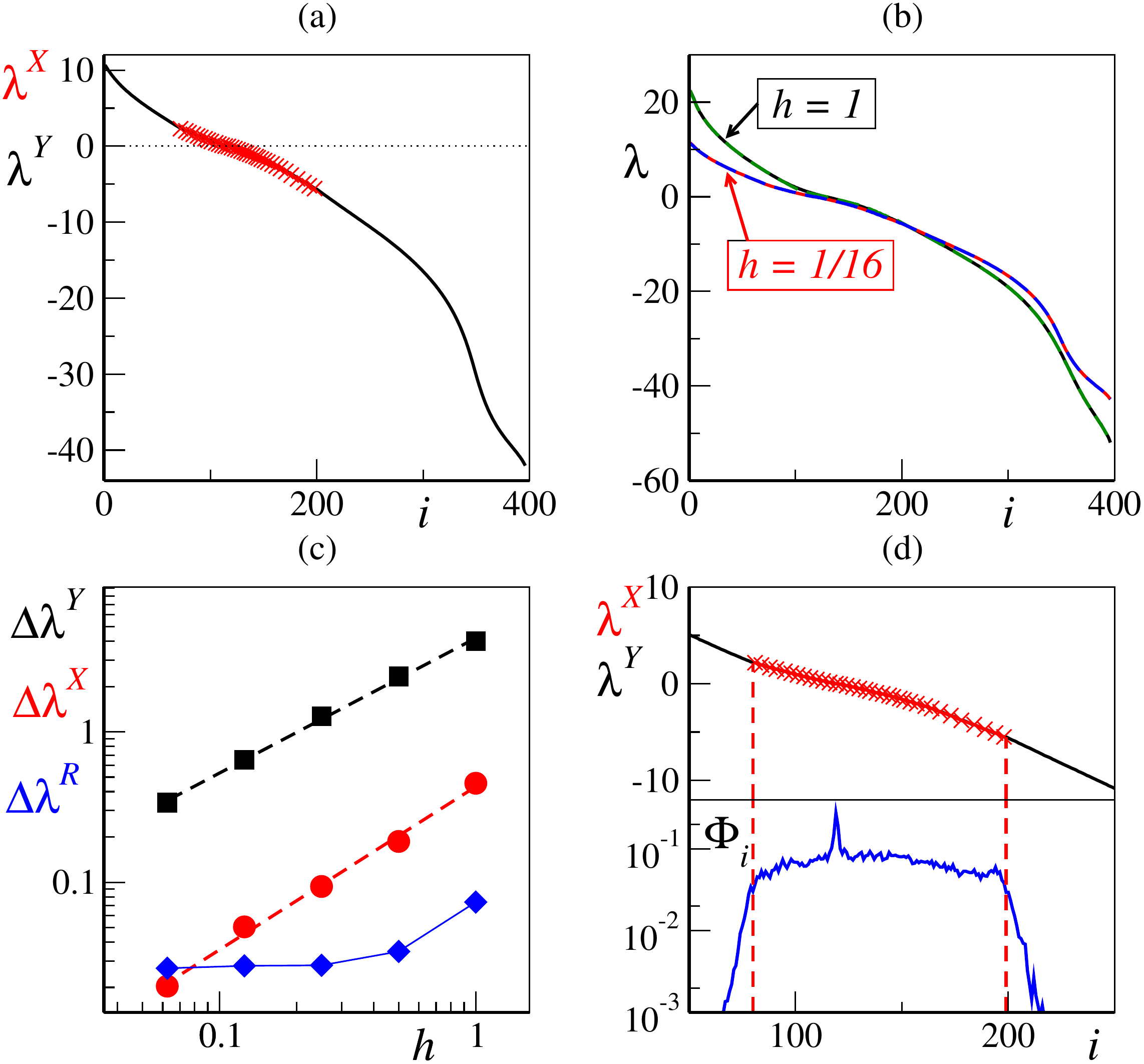}
	\caption{(a) Lyapunov sepctra for the uncoupled ($h=0$) case decomposed in its slow ($\lambda^X$, red crosses) and fast ($\lambda^Y$, full black line) parts. (b) Full Lyapunov spectra (full lines) superimposed with spectra reconstructed from their restricted (see text) counterparts (dashed lines) for $h=1$ and $h=1/16$.
(c) Root mean squared difference between the finite $h$ restricted spectra and the completely uncoupled fast and slow spectra, for both fast (black squares) and slow (red circles) dynamics. Power law fits (dashed lines) return slopes close to unity (respectively $\approx 0.9$ and 
$\approx 1.1$) suggesting a simple linear growth with the coupling $h$. Blue diamonds refer to the root mean squared difference $\Delta \lambda^R$ between the full LS and the reconstructed ones in the restricted tangent space approximation. 
Both axes are represented in a doubly-logarithmic scale.    
(d) Upper panel: Zoom on the central part ($i \in [50, 250]$ of the restricted spectra for $h=1/8$ with the slow ($\lambda^X$, red crosses) and fast  ($\lambda^Y$, full black line) components differently marked. Lower panel: Same zoom of the $h=1/8$ projection norm as computed from the fully coupled dynamics. The vertical red dashed lines mark the upper ($i_L$) and lower ($i_R$) boundaries of the superposition region as reported in table 1. 
In all panels we have fixed $b=10$, $K=36$ and $J=10$.} 
	\label{Fig8}
\end{figure}

We begin by discussing  the pedagogical example of the uncoupled limit ($h=0$).
In this case, the $X$ and $Y$ subsystems evolve, by definition, independently,  and one can separately determine $K$ LEs
associated with the slow variables and $KJ$ exponents associated with the fast variables.
The full spectrum can be thereby reconstructed by combining the two distinct spectra into a single
one. The result is illustrated in Fig.~\ref{Fig8}a, where the red crosses correspond to the
$X$ LEs.  Note that the same area is also spanned by the central part of the fast-variable spectrum.
The region covered by the slow LEs, where the instability rates of the two uncoupled systems have the same magnitude, roughly corresponds to the 
location of the slow bundle in the coupled-model CLVs spectrum. 
This suggests that the origin of the slow bundle can be traced back to a sort of {\it resonance} 
between the slow variables and a suitable subset of the fast ones.

Note also that, in the absence of coupling, the Jacobian has a block diagonal structure, with the CLVs either belonging to the $X$, or $Y$ subspaces.
The projection $\Phi_i$, therefore is strictly equal to either 0 or 1, depending on the vector type, and can be used 
to distinguish the two types of vectors when the full set of (uncoupled) equations is integrated simultaneously.

We now proceed to discuss the coupled case. When the coupling is switched on, it has a double effect:
(i) it modifies the overall dynamics, i.e. the evolution in
phase space, Eq.~(\ref{Eq::Original}); (ii) it affects directly the tangent-space evolution,  
Eq.~(\ref{Eq::Tangent}), 
destroying the block diagonal structure of the uncoupled Jacobian. This, in turn, prevents one from identifying single LEs with either the slow or the fast dynamics.
In order to be able to distinguish the two contributions also in the $h>0$ case, we study an intermediate setup characterized by a full coupling
in real space, but removing it from the tangent-space dynamics. In practice, we simulate the full nonlinear model (\ref{Eq::Original}), 
and use the resulting trajectories to ``force'' an {\it uncoupled} tangent space dynamics, that is  
\begin{subequations}
\begin{align}
\label{linearX1}
&\delta\dot{X}_k = \delta X_{k-1} \; (X_{k+1}-X_{k-2}) + X_{k-1} \; (\delta X_{k+1}- \delta X_{k-2})-\delta X_k \\
&\label{linearY1}
\delta\dot{Y}_{k,j} = cb \big[ \delta Y_{k,j+1} \; (Y_{k,j-1}-Y_{k,j+2}) + Y_{k,j+1} \; (\delta Y_{k,j-1}- \delta Y_{k,j+2}) \big]-c \delta Y_{k,j}  \,,
\end{align}
\label{Eq::TangentR}
\end{subequations}
where the coupling terms, proportional to $h$ in Eqs. (\ref{Eq::Tangent}), have been ignored.
This way the Jacobian matrix is still block diagonal.

Thanks to this approximation, we can define two {\it restricted} spectra $\lambda_k^X$ and $\lambda_j^Y$ 
for the slow and fast variables, respectively, and thereby recombine them into a single spectrum, 
by ordering the exponents from the largest to the most negative one.
A comparison between the resulting reconstructed spectra and the full ones (with coupling acting both in real and tangent space) 
shows an excellent agreement, at least in the range $h \in (0,1)$. Two examples, for $h=1$ and $h=1/16$ are given in Fig.~\ref{Fig8}b,
while the dependence of their root mean squared difference $\Delta \lambda^R$ on $h$ is reported in Fig.~\ref{Fig8}c (blue diamonds).
It is not however clear to what extent this is a general property of high-dimensional dynamics:
we are not aware of similar analyses made in high-dimensional models.

The modifications induced by real space coupling are more substantial. 
They can be quantified by computing the root-mean-square differences 
\begin{eqnarray}
\Delta \lambda^X (h)&=& \sqrt{\frac{1}{K}\sum_{k=1}^K \left[\lambda_k^X(h) - \lambda_k^X(0)\right]^2}\\
\Delta \lambda^Y (h)&=& \sqrt{\frac{1}{K J}\sum_{j=1}^{KJ} \left[\lambda_j^Y(h) - \lambda_j^Y(0)\right]^2}
\; ,\nonumber
\label{rmsq}
\end{eqnarray}
which measure the average variation of the restricted spectra upon increasing the coupling. 
Numerical simulations, reported in Fig.~\ref{Fig8}c, show that both $\Delta \lambda^X$ and 
$\Delta \lambda^Y$ increase approximately linearly with $h$, the main difference being that $\lambda_k^X(h)$ decrease (in absolute value) when $h$ is increased, while the opposite holds for $\lambda_j^Y(h)$.
In Fig.~\ref{Fig8}c, we can also appreciate that both $\Delta \lambda^X$ and $\Delta \lambda^Y$ are significantly
larger than $\Delta \lambda^R$, showing that tangent-space coupling is not relevant 
for the estimate of the LEs. On the other hand, this is not expected to be true for the CLVs,
which are local objects, defined at each attractor point.
CLVs associated with the restricted LEs are, by definition, confined either to the
$X$ or the $Y$ subspace, that their $X$-projection $\Phi_i$ is again either equal to 0 or 1.
The analysis carried out in the previous section for the fully coupled dynamics shows instead that 
the average projection $\Phi_i$ of any vector within the slow bundle is significantly 
different from both 0 and 1 and substantially constant over the entire central band.
This means that the orientation of the CLVs is very sensitive to the coupling itself. 

The main mechanism responsible for the reshuffling of the CLV orientation is the (multifractal) 
fluctuations of finite-time LEs ~\cite{PikovskyPoliti_LE}.
Fluctuations are the unavoidable consequence of the different degree of stability experienced
in different regions of the phase-space and they occur in both strictly hyperbolic and
non-hyperbolic dynamical systems, although they are typically much larger in the latter
context. Fluctuations may be so large as to bridge the gap between distinct LEs, 
which results in a lack of domination of the Oseledets splitting \cite{Pugh04, Viana05}
and in the sporadic occurrence of near tangencies between pairs of different 
CLVs~\cite{Kaz09,Kaz11}\footnote{Perfect tangencies may occur, but only for a set of zero-measure 
initial conditions, such as the homoclinic tangencies in low-dimensional chaos.}.
Fluctuations are also responsible for the so-called 
{\it coupling sensitivity}~\cite{Daido,PikovskyPoliti_LE}: strictly degenerate LEs in uncoupled
systems may separate by an amount of order $1/|\ln \varepsilon|$,
where $\varepsilon$ is the (small) amplitude of the coupling strength.

Let us be more quantitative and introduce the finite-time Lyapunov exponents $\gamma_i (t)$,
computed from the average expansion rate over a window of length $\tau_w$,
\begin{equation}
\gamma_i (t) = \frac{1}{\tau_w} \ln ||{\bf M}({\bf x}_t, \tau_w) {\bf v}_i(t) || \; ,
\label{FTLE}
\end{equation}
where ${\bf M}({\bf x}_t, \tau_w)$ is the propagator (\ref{Eq::M}) for the tangent-space 
evolution over time $\tau_w$, while the CLVs ${\bf v}_i$ is normalized to unity. Their asymptotic time average obviously coincides with the corresponding LEs, 
$\langle \gamma_i (t) \rangle_t \equiv \lambda_i$.

\begin{figure}[t!] 
  \centering
  \includegraphics[width=0.85\linewidth]{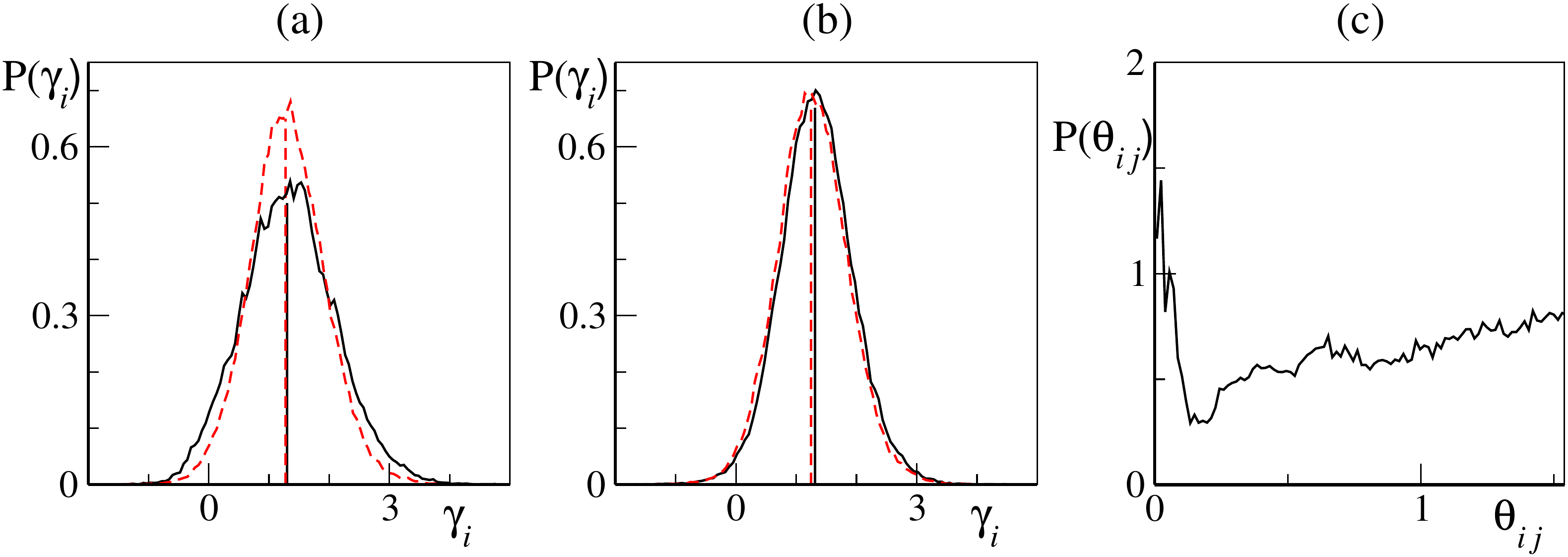} 
\vspace{0.5cm} 
\caption{(a)-(b) Probability distribution of the finite-time LEs (\ref{FTLE}) for two nearby LEs belonging to the slow bundle.
We show results for the 92nd (full black line) and 93rd (dashed red line) LEs which, in the absence of tangent space coupling (the {\it restricted} setup, see text), belong respectively to the $\lambda^X$ and $\lambda^Y$ restricted spectra. 
(a) Finite-time LEs fluctuations in the restricted case. The vertical lines mark the mean values $\lambda_{92}^X=1.30$ and  $\lambda_{93}^Y=1.275$ (indexes refers to the full spectrum position. They correspond to the 5th and 88th (respectively) LE in the $X$ and $Y$ restricted spectra).
(b) Finite-time LEs fluctuations in the fully coupled case. The vertical lines mark the mean values $\lambda_{92}=1.31$ and  $\lambda_{93}=1.25$.
(c) Probability distribution of the angle $\theta_{ij}$ (see Eq. \ref{angle}) between the two corresponding CLVs in the fully coupled case.
All simulations have been performed for $K=36$, $J=10$, $b=10$ and $h=1/16$. We have fixed $K=36$ and $J=10$ in both panels.}
  \label{Fig::FTLE}
\end{figure}

We are interested in the probability distribution $P(\gamma_i)$ of $\gamma_i$, obtained by evolving a 
long trajectory. For short times, $\gamma_i$ fluctuations significantly depend on the coordinates
used to parametrize the dynamics, but upon increasing $\tau_w$, such a variability is progressively lost
and the width of $P(\gamma)$ scales as $1/\sqrt{\tau_w}$ as prescribed by the multifractal 
formalism~\cite{PikovskyPoliti_LE}.
In the following, we have set $\tau_w=0.5$, after having verified that it is long enough.
Here, for illustrative purposes, we have selected two vectors which, in the absence of tangent-space coupling, are of $X$ and $Y$ type, 
respectively\footnote{Given the two restricted spectra $\lambda_k^X$ and $\lambda_j^Y$, they are combined into a single set of ordered LEs and labeled according to the index $i$. Depending whether the $i$-th exponent belongs to the $X$ or $Y$ restricted spectra, we conclude that the corresponding $i$-th CLV in the fully coupled spectrum is of $X$ or $Y$ type. }.
From Fig.~\ref{Fig::FTLE}a, it is clear that the amplitude of the fluctuations largely exceeds 
the difference between the corresponding mean values (i.e., the asymptotic LEs, see the vertical straight lines) 
and that the same holds true after restoring the coupling in tangent space (Fig.~\ref{Fig::FTLE}b).

Consistently, in Fig.~\ref{Fig::FTLE}c we show that the corresponding CLVs are characterized by non negligible near tangencies: 
The probability distribution of the relative angle
\begin{equation}
\theta_{i,j} (t) = \arccos \left[ {\bf v}_i(t) \cdot {\bf v}_j(t) \right]\,,
\label{angle}
\end{equation}
indeed exhibits a peak near 0.
 
We have verified this to be the generic behavior, as expected due to the non hyperbolic nature of the L96 model.
The near tangencies between different CLVs within the slow bundle provide strong numerical evidence of the ``mixing'' between 
slow and fast degrees of freedom, and are perfectly consistent with the non-negligible projection 
on the slow $X$-subspaces of all vectors in the slow bundle. In fact, in the presence of two similarly 
unstable directions, the corresponding CLVs tend to wander in a (fluctuating) 
two-dimensional subspace selecting their current direction on the basis of the relative degree of instability.
It is therefore natural to expect that, in the presence of strong fluctuations, an $X$-type vector 
(in the uncoupled limit) temporarily aligns along the $Y$ directions and viceversa, 
thereby giving rise to a projection pattern such as the one seen in the central region of the CLVs spectrum,
where all vectors have a nonnegligible average projection over the $X$ degrees of freedom.

The intermittent nature of the instantaneous $X$-projection $\phi_i(t)$ discussed in Sec.~\ref{proj} further
validates this picture: it is the result of the large fluctuations exhibited by finite-time LEs, which are, in turn,
associated with changes of directions when the CLV comes close to tangencies.
Furthermore, the large ratio between
the amplitude of the fluctuations and the separation between consecutive LEs suggests that
this exchange of directions may extend beyond nearest-neighbours along the spectrum.
We conjecture that the relatively smooth boundary of the central band
is precisely a manifestation of this sort of extended interaction.
 
Finally, we return to the restricted LEs to see whether -- as implied by the above conjecture -- their knowledge can help to
identify the slow bundle boundaries. In practice, we have first identified the borders of the region covered by both slow LEs. 
They are given by the indices (within the reconstructed spectrum) of the largest and smallest restricted 
slow LE, labelled respectively as $i_L$ and $i_R$. 
They are reported in Table 1, together with the corresponding value of the restricted Lyapunov exponent, for different coupling values.
The agreement with the actual boundaries of the slow bundle -- as revealed by a visual inspection of the projection patterns $\Phi_i$ -- is actually pretty good (see, e.g., 
Fig.~\ref{Fig8}d for $h=1/8$).
\begin{table}[h!]
  \begin{center}
 \caption{$X$-restricted largest  ($\lambda^X_1$) and smallest ($\lambda^X_K$) LE with the corresponding full spectrum indexes 
($i_L$ and $i_R$). All data refer to $b=10$, $K=36$ and $J=10$.}    
 \begin{tabular}{c||c|c|c|c} 
      $h$ & $\lambda^X_1$ & $\lambda^X_K$ & $i_L$ & $i_R$\\
      \hline
      1 & 1.33 & -4.57 &  104 & 192 \\
      1/2 & 1.89 & -5.18 & 93 & 197 \\
      1/4 & 2.13 & -5.42 & 85 & 198\\
      1/8 & 2.23 & -5.53  & 80 & 199\\
      1/16 & 2.29 & -5.58 & 77 & 199\\
\end{tabular}
  \end{center}
  \label{table1}
\end{table}
Altogether, our analysis suggests that coupling 
in real space induces a sort of ``short-range'' interaction within tangent space: 
each LE (and the corresponding CLV) tends to affect and be affected by exponents with similar magnitude and thereby
characterizing a similar degree of instability, in a sort of resonance phenomenon.

Note finally that the spectral band where the slow and fast restricted Lyapunov spectra superimpose covers all the slow restricted LEs and a {\it finite} fraction of the fast ones, thus providing a justification of the phenomenological scaling law (\ref{Eq::scaling}).

\section{Finite perturbations}
\label{fs}

So far we have studied the geometry of the L96 model, dealing exclusively with infinitesimal perturbations. 
A legitimate question is whether we can learn something more, by looking at {\it finite} perturbations.

Finite-size analysis has been implemented in the L96 model since its introduction \cite{Lorenz96} and it has been formalized with the definition of the so-called Finite Size Lyapunov Exponents (FSLEs) \cite{FSLE}. In a nutshell, the rationale for introducing FSLEs is -- as already recognized by Lorenz -- that in nonlinear systems the response to finite perturbations may strongly depend on the observation scale. Dropping the limit of vanishing perturbations, of course, weakens the level of mathematical rigor of the infinitesimal Lyapunov analysis, but it nevertheless allows for a meaningful study of the underlying instabilities.

Here, we follow the excellent review \cite{Cencini}, where applications to L96 were also discussed.
Given a generic trajectory ${\bf x}(t)$, the idea is to define a series of thresholds $\delta_n = \delta_0 \sigma^n$, with $\sigma>1$, and to measure the times $\tau(\delta_n)$ needed by the norm of a finite perturbation $\Delta {\bf x}(t)={\bf x}'(t)-{\bf x}(t)$ to grow from the amplitude $\delta_n$ to $\delta_{n+1}$. The FSLE $\Lambda(\delta_n)$ is then defined as
\begin{equation}
\Lambda(\delta_n) = \frac{\ln \sigma}{\langle \tau(\delta_n) \rangle}\,,
\label{FSLE}
\end{equation}
where $\langle \cdot \rangle$ denotes an average over many realizations of the perturbation.
In practice, one starts at time $t_0$ with a finite perturbation $\left\lVert \Delta {\bf x}(t_0) \right\rVert \ll \delta_0$ to ensure a correct
alignment (along the most expanding direction) by the time the perturbation amplitude reaches the first threshold $\delta_0$. 
Subsequentely, both trajectories ${\bf x}$ and ${\bf x}'$ are followed, registering the crossing times of all 
$\delta_n$ thresholds. By repeating this procedure many times, one is able to estimate the FSLEs for all amplitudes $\delta_n$ 
via Eq.~(\ref{FSLE}).
The FSLE in principle depends on the norm used to define the size of the perturbation~\cite{Cencini}. 
However, by construction, for vanishing perturbations, the FSLE should coincide with the largest LE regardless of the norm
\begin{equation}
\lim_{\delta \to 0} \Lambda(\delta) = \lambda_1 \; .
\label{FSLE2}
\end{equation}
In Ref.~\cite{Cencini}, FSLEs have been applied to analyse the L96 model (see also \cite{Boffetta98} for an earlier study). In a slightly different setup (no fast-variable forcing and a larger scale separation $b$), it was shown that the FSLE is characterized by two different plateaus: (i) a small-$\delta$ one, essentially associated with the instability of the fast, convective, degrees of freedom and roughly equivalent to the largest LE, $\Lambda(\delta) \approx \lambda_1$; 
(ii) a ``large"-$\delta$ plateau that was associated with the intrinsic instability $\Lambda_s$ on the slow larger scales. 
Interestingly, it was observed that also the height of the second plateau seems to be roughly norm independent. 
This observation led to the conjecture that the nonlinear evolution of large perturbations may be controlled by the linear dynamics of an effective lower dimensional system, capturing the essence of the slow-variable dynamics \cite{Cencini}.

In this section we repeat this analysis in our setup, comparing the behavior of the FSLE with the analysis of the 
tangent-space slow bundle.  In the following we use our standard parameters ($K=36$, $J=10$, $b=10$), 
setting $\sigma = \sqrt{2}$, $\delta_0=10^{-3}$ and averaging the crossing-times over $10^3$ realizations. 
For each realization, the initial finite perturbation $(\Delta X_1, \ldots, \Delta X_K, \Delta Y_{1,1}, \Delta Y_{K,J})$ is chosen at random with an initial amplitude of $10^{-5}$. 

As we expect the FSLE to depend on the norm, we have decided to transform this weakness into an advantage,
by studying the behavior of an entire family of Euclidean norms and thereby extracting useful information from
the dependence on the chosen norm.
More precisely, we introduce the $\gamma$-dependent norm
\begin{equation}
\left\lVert\cdot\right\rVert_\gamma = \sqrt{\sum_k X_k^2 + \gamma \sum_{k,j} Y_{k,j}^2}\, ,
\label{g-norm}
\end{equation}
which, for $\gamma=1$, coincides with the standard Euclidean norm. We consider $\gamma \in [0, 1]$, a choice which
allows exploring a broad range of weights of the slow-variables.
\begin{figure}[!t]
\centering
 \includegraphics[width=0.7\textwidth]{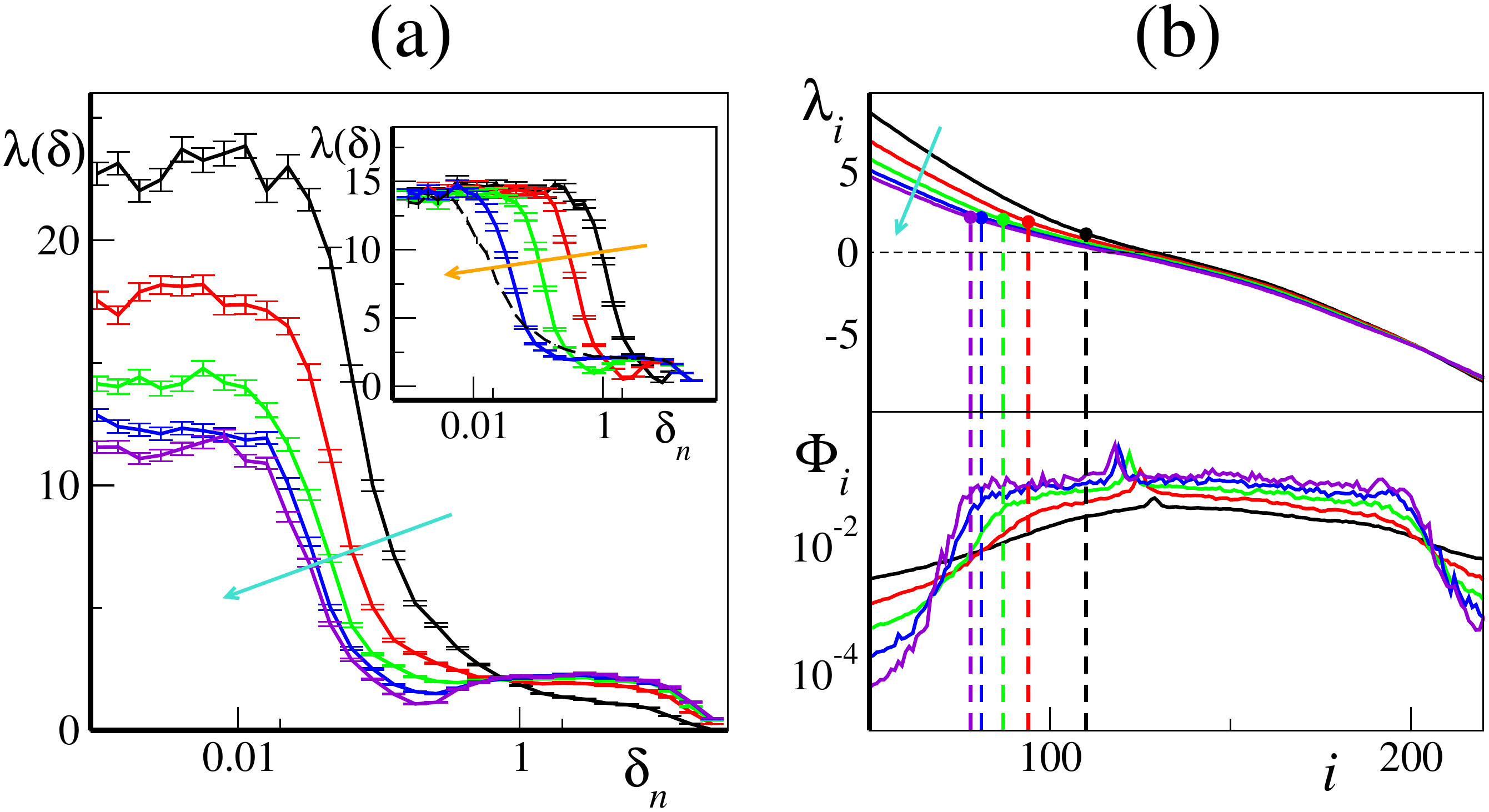}
	\caption{FSLE analysis. (a) FSLEs for different coupling constants, $h=1$ (black line), $h=1/2$ (red line), $h=1/4$ (green line), $h=1/8$ (blue line), 
$h=1/16$ (red line) -- decreasing along the cyan arrow -- vs. the finite size amplitudes $\delta_n$ for the $\gamma=10^{-3}$ norm. Inset: FSLEs for $h=1/4$ and different $\gamma$-norms ($\gamma=1,\,10^{-1}\,10^{-2}\,10^{-3}$, decreasing along the orange arrow). The error bars measure one standard error. The dashed black line marks the FSLEs for the limiting norm $\gamma=0$.
Note the logarithmic scale for the abscissa of both graphs. (b) Details of the LE spectrum (top panel) and of the $\Phi_i$ average projection norm patterns (lower panel) as in Fig.~\ref{Fig3}b. The full dots and the vertical dashed line mark the value and the location of the LEs $\lambda_{i_S}$ as identified in Table 2 (see main text for more details). 
Different coupling constant values are color coded as in panel (a), with $h$ decreasing along the cyan arrow in the top panel.}
	\label{Fig6}
\end{figure}
In the inset of Fig.~\ref{Fig6}a, we see that the main effect of changing the norm is a variation in the 
length of the two plateaus: upon decreasing $\gamma$, the first plateau shrinks, 
leaving space for a longer second plateau. The height of the two plateaus is largely $\gamma$-independent.
This behavior can be qualitatively understood as follows. At early times, all components of the perturbation 
grow according to the maximum LE, which we know from the previous analysis to be 
mostly controlled by the dynamics of the fast $Y$ variables.
As time goes on, the perturbations of the $Y$ variables start saturating, while those of the 
slow variables keep growing, at a pace, however, controlled by their (weaker) intrinsic instability.
Upon decreasing $\gamma$, the relative weight of the less unstable, slow variables increases. 
However, there is a limit: even when $\gamma=0$, the growth rate of the $X$-perturbations is initially
controlled by the fast variable. The range of this initial, fully linear, regime, depends on the
initial amplitude of the fast components: this limit corresponds to the dashed curve into the inset
of Fig.~\ref{Fig6}a.

The FSLEs obtained for different coupling parameters $h$ are shown in Fig.~\ref{Fig6}a, 
all for $\gamma=10^{-3}$.  Two plateaus are clearly visible in all cases. 
The first one coincides with the maximum LE of the whole system, as per Eq.~(\ref{FSLE2}). 
The second one approximately extends over a range of one order of magnitude
(at large scales the plateau is obviousy limited by the attractor size). 
Its height corresponds to the characteristic instability $\Lambda_s (h)$ associated with the 
effective dynamics of the slow variables, as conjectured in \cite{Cencini}. 

For each coupling $h$, we estimated the corresponding $\Lambda_s (h)$ and identified 
the closest LE $\lambda_{i_S}$ and its index $i_S$ in the Lyapunov spectra.
The results of this procedure are summarized in Table 2 and compared with the results of the 
restricted analysis obtained in section~\ref{sec4}.

\begin{table}[h!]
  \begin{center}
 \caption{Estimated slow dynamics finite size instability $\Lambda_s (h)$ compared to the closest LE and its index $i_S$ for different coupling constants. 
     $\Lambda_s$ has been typically averaged over the range  $\delta \in [1,10]$. Results of the restricted analysis carried on in section \ref{sec4} are reported from table 1 for comparison.
All data has been obtained with $b=10$, $K=36$ and $J=10$.}    
 \begin{tabular}{c||c|c|c|c|c} 
      $h$ & $\Lambda_s$ & $\lambda_{i_S}$ & $i_S$ & $\lambda^X_1$ & $i_L$ \\
      \hline
      1 & 1.2(3) &1.15 &  110(6) & 1.33 & 104\\
      1/2 & 1.86(11) & 1.90 & 94(2) & 1.89 & 93 \\
      1/4 & 2.07(8) & 2.04 & 87(1) & 2.13 & 85 \\
      1/8 &2.16(11) & 2.15 & 81(1) & 2.23 & 80\\
      1/16 &2.2(1) & 2.18 & 78(1)  & 2.29 & 77\\
\end{tabular}
  \end{center}
  \label{table2}
\end{table}
In practice, by interpreting the height of each plateau as a suitable instability rate within the full Lyapunov spectrum, one
can thereby extract the corresponding index $i_S$ for each value of the coupling constant. 
In Fig.~\ref{Fig6}b we have marked these indices with dashed vertical lines (upper panel)
and compared with the slow bundle projection patterns (lower panel).
Interestingly, these values seem to provide a reasonable estimate of the leftmost boundary 
of the central band which defines the slow bundle in tangent space.
Essentially, $i_s$ coincides fairly well with the left ``shoulder'' 
where $\Phi_i$ starts to drop towards negligible values for decreasing $i$.
Note however, that for larger values of $h$, the second plateau becomes less sharply defined, 
up to the case $h=1$, where it is practically impossible to define a threshold. 
Correspondingly, also the boundaries of the slow bundle in tangent space, 
as defined by inspection of $\Phi_i$, become less well defined. 

Altogether, the slow-variable (large-scale) instability $\Lambda_s$ emerging from the finite-size analysis 
roughly coincides with the upper boundary of the slow bundle (i.e. the LEs associated with the 
CLVs with a relevant projection on the $X_k$ variables).
Following the analysis of the restricted spectra carried out in the previous section, $\Lambda_s$ 
is also close to the first restricted LE associated with the $X$ subspace $\lambda^X_1$, as shown in Table 2.
The parameter $\gamma$ proves to be useful in improving the accuracy of the two plateaus exhibited
by the FSLEs.

It is remarkable that the analysis of a single pair of trajectories allows extracting
information about (at least) two different Lyapunov exponents.
We conjecture that the linearly-controlled growth of small, finite perturbations stops
as soon as the fast components saturate because of nonlinearities.
Afterwards, fast variables act as a sort of noise on the slow ones, whose
dynamics is still in the linear regime. 
Finally, in view of the above mentioned closeness between restricted and fully coupled LS, it is reasonable to conjecture that, since coupling does not play a crucial role in tangent space, 
the growth rate corresponds, in this second regime, to the maximal LE of the slow variables, as
indeed observed.

Our result supports an earlier conjecture of Ref.~\cite{Cencini} concerning the existence 
of an effective lower dimensional dynamics capturing the slow-variable behavior.

\section{Conclusion} 
Our analysis of the tangent-space structure of the L96 model has identified a {\it slow bundle} 
within the full tangent space. It is composed of the set of covariant Lyapunov vectors characterized 
by a non-negligible projection over the slow degrees of freedom. Vectors in this set are associated 
to the smallest (in absolute value) LEs, and thus to the longest timescales. 
We have verified that the number of such vectors increases linearly with the total number of 
degrees of freedom, so that the slow bundle dimension is an extensive quantity.

\begin{figure}[!t]
\centering
 \includegraphics[width=0.7\textwidth]{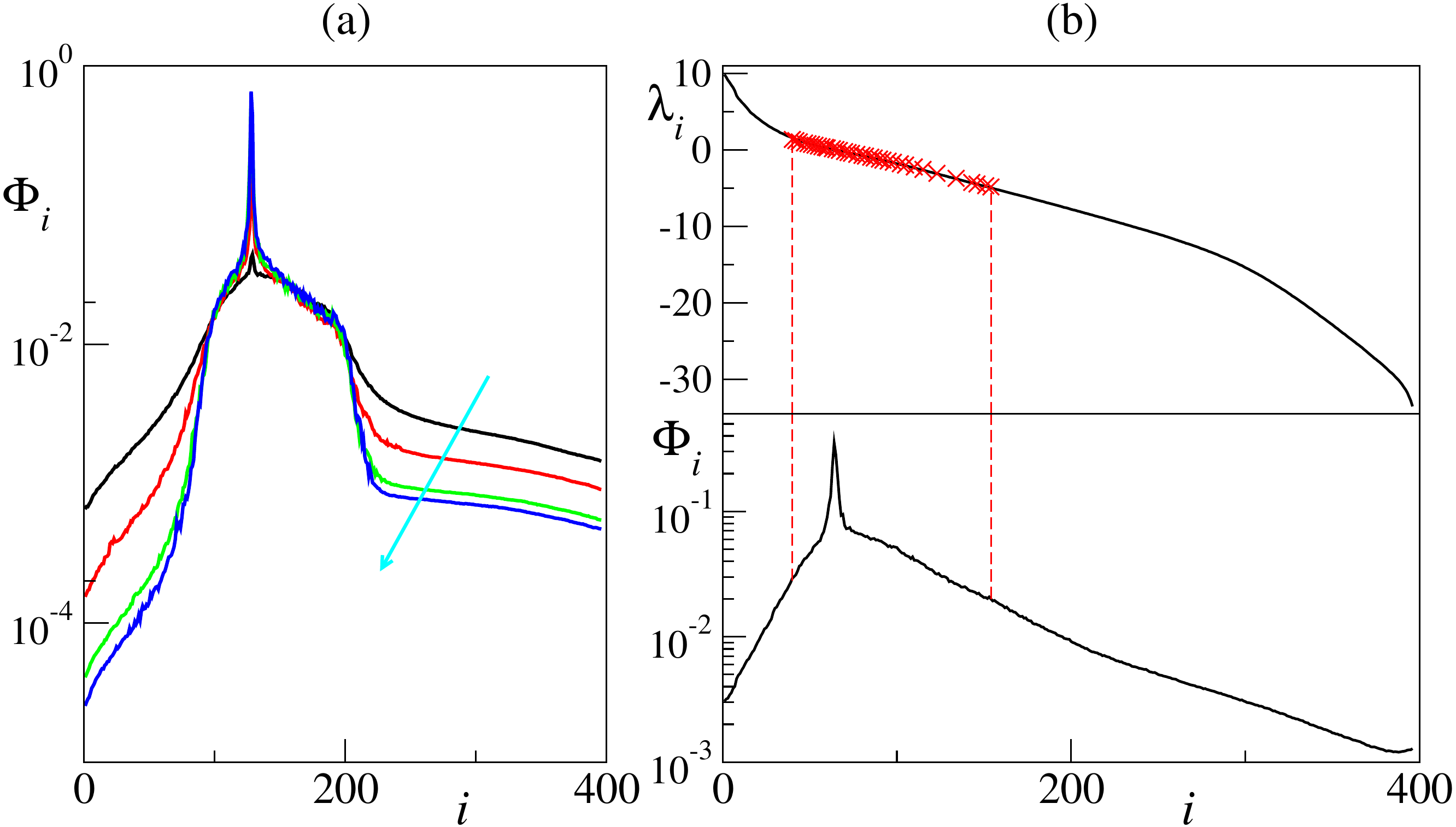}
\vspace{0.2cm}
	\caption{a) X-projection patterns $\Phi_i$ for Lorenz '96 with fast variable forcing ($F_f=6$), strong coupling $h=1$ and increasing (along the direction of the cyan arrow) values of amplitude separation, $b=10, 20, 40, 50$. 
(b) Lyapunov Spectrum (top panel) and X-projection patterns (lower panel) for Lorenz '96 with no fast variable forcing ($F_f=0$) and "standard" parameter values, $h=1$, $b=c=10$. The red crosses mark the values of the $X$-restricted spectrum (see Section 4 for more details).
System size is $K=36$ and $J=10$ in both panels.}
	\label{FigLast}
\end{figure}

The upper and lower boundaries of the slow bundle are better defined for a weak coupling $h$.
However, the rescaled formulation of L96 (see Eq.~(\ref{RescaledX})) shows that the effective
upward coupling (from the fast to the slow variables) is $h f = hJc/b^2$, thereby suggesting that
an increase of the amplitude separation $b$ can  increase the sharpness of the slow-bundle 
boundaries even for large $h$. As reported in Fig.~\ref{FigLast}a,
numerical simulations with $h=1$ and increasing values of $b$, 
actually confirm this intuition, indicating that a slow bundle can be clearly defined also in 
the strong coupling limit, provided that the slow- and fast-scale amplitudes are sufficiently separated.

In order to clarify the origin of the slow bundle, we have introduced the notion of restricted Lyapunov spectra 
and argued that the central region, where the CLVs retain a significative projection over both slow and fast 
variables, corresponds to the range where the restricted spectra overlap with one another. 
In this region, fluctuations of the finite-time LEs much larger than the typical separation between consecutive 
LEs lead inevitably to frequent ``near tangencies'' between CLVs, thereby ``mixing'' slow and
fast degrees of freedom into a non-trivial set of vectors which carry information on both set of variables.

Besides, we have found that coupling in tangent space weakly influences the actual LEs, provided that 
it is accounted for in real space. This is one of the reasons for the finite-size analysis being able
to give information about the instability of the slow bundle (i.e. the correspondence between 
the second plateau displayed by the FSLE and the upper boundary of the slow bundle).
Further investigations are necessary to put our consideration on a firmer ground.

So far, we have discussed the slow bundle in a setup where the fast degrees of freedom are forced by a strong external drive $F_f$, so that the fast dynamics is intrinsically chaotic even in the absence of coupling. 
One might wonder how general these results are and, in particular, how they could be extended to the traditional 
L96 setup, with no forcing of the fast variables \cite{Lorenz96}. 
When $F_f=0$, in the zero-coupling limit, the fast dynamics is dominated by dissipation with no chaotic features.
However, it is easy to verify that for a sufficiently strong coupling, 
the fluctuations of the slow variables induce a chaotic dynamics on the fast ones as well, 
so that the ``classical'' setup resembles the forced one analyzed in this paper. 
While we have not performed an accurate and thorough study, preliminary simulations 
indicate that the signature of a slow bundle can be found also in the classical setup
for sufficiently strong coupling. 
In particular, as reported in Fig.~\ref{FigLast}b for $h=1$, one can see that the region of 
non-negligible $X$-projections of the CLVs again 
coincides with the superposition region of the slow and fast restricted spectra.

As already mentioned, the slow bundle is identified as the set of CLVs with a non negligible projection on the slow degrees of 
freedom. One might argue that the average projection $\Phi$ on the $X$ subspace decreases with $J$,
being at best of order $1/J$, i.e. the fraction of slow degrees of freedom. However, what matters is not
the actual value of $\Phi$ but rather the ratio between the height of the plateau and that of the
underlying background. The scaling analysis reported in Fig.~\ref{Fig4} shows that this ratio stays finite
while increasing the number of fast variables.

Altogether, we conjecture that: (i) the fast stable directions lying beyond the slow bundle central region are basically 
slaved degrees of freedom, which do not contribute to the overall dynamical complexity;
(ii) the fast unstable directions act as a noise generator for the $Y$ degrees of freedom (their
projection on the $X$ variables being negligible, they do not {\it talk} directly with the slow variables).
Therefore, it is natural to conjecture that the two subsystems mutually
interact only through the slow bundle instabilities, so that suitably aligned perturbations of the fast variables can affect the slow variables and
vice versa, see also \cite{Vannitsem2016}.
While this is a mere conjecture, to be explored in future works, it suggests that (some) covariant vectors
could be profitably applied to ensemble forecasting and data assimilation in weather and climate models. 
In particular, it has been already shown that restricting variational data assimilation to the full unstable subspace can increase the forecasting efficiency \cite{Trevisan2004, Trevisan2010}. 
In the future, we would like to explore whether a data assimilation scheme restricted to 
the slow bundle only (which {\it do not} encompass the entire unstable space) 
can lead to further improvements in forecasting.

The mechanism discussed in Sec.~\ref{sec4}, relying on the overlap of the two restricted Lyapunov spectra, should be common in nonlinear multiscale systems; therefore we believe our findings to be fairly generic. In the future, it will be interesting to extend the present analysis of Lyapunov exponents and covariant Lyapunov vectors to models with multiple scales and/or higher complexity and relevance, such as the coupled atmosphere-ocean model MAOOAM \cite{Cruz2016} or simplified multilayer models of the atmosphere such as PUMA \cite{Fraedrich2005}) or SPEEDY \cite{Molteni2003}, thus going beyond the LEs studies of \cite{DeCruz(2018)} to include the full tangent space geometry.

\section*{Acknowledgement}
FG warmly thanks M. Cencini for truly invaluable early discussions.
We acknowledge support from EU Marie Sklodowska-Curie ITN Grant No. 642563 (COSMOS).
MC acknowledges financial support from the Scottish Universities Physics Alliance (SUPA), as well as Sebastian Schubert and the Meteorological institute of the University of Hamburg for the warm welcome and the stimulating discussions.
VL acknowledges the support received from the DFG Sfb/Transregion TRR181 project and the EU Horizon 2020 project Blue Action.


\bibliographystyle{apalike}

\begin{thebibliography}{99}

\bibitem[Abramov and Majda(2007)]{Abramov2008} Abramov, R. V., and Majda, A.: New approximations and tests of linear fluctuation-response for chaotic nonlinear forced-dissipative dynamical systems. Journal of Nonlinear Science, 18:303-341, 2008. 10.1007/s00332-007-9011-9, 2007.

\bibitem[Aurell et al.(1997)]{FSLE}
  Aurell, E., Boffetta, G., Crisanti, A., Paladin, G., and  Vulpiani, A.: Predictability in the large: an extension of the concept of Lyapunov exponent, J. Phys. A, 30 1, 1997.

\bibitem[Benettin et al..(1980)]{Benettin80}
Benettin, G., Galgani, L., Giorgilli, A., and Strelcyn, J. M.: Lyapunov characteristic exponents for smooth dynamical systems and for Hamiltonian systems; a method for computing all of them. Part 1: Theory, Meccanica 15 9, 1980.
Benettin, G., Galgani, L., Giorgilli, A., and Strelcyn, J. M.: Lyapunov Characteristic Exponents for smooth dynamical systems and for hamiltonian systems; A method for computing all of them. Part 2: Numerical application, Meccanica 15 21, 1980.

\bibitem[Berner et al.(2017)]{Berner2017}
Berner, J. et al.: Stochastic parametrization: Toward a new view of weather and climate models. Bull. Am. Meteor. Soc. 98, 565, 2017.

\bibitem[Blender et al.(2013)]{Blender2013} 
Blender, R., Lucarini, V., and Wouters, J.: Avalanches, breathers, and flow reversal in a continuous Lorenz-96 model, Phys. Rev. E 88, 013201, 2013.

\bibitem[Bochi and Viana(2005)]{Viana05}
 Bochi, J. and  Viana, M.: The Lyapunov exponents of generic volume-preserving and symplectic maps, Ann. Math. 161, 1423, 2005.

\bibitem[Boffetta et al.(1998)]{Boffetta98} Boffetta, G., Giuliani, P., Paladin, G., and Vulpiani, A., J.: An Extension of the Lyapunov Analysis for the Predictability Problem, Atmos. Sci. 55 3409, 1998.

\bibitem[Cencini and Vulpiani(2013)]{Cencini} Cencini, M., \& Vulpiani, A.: Finite size Lyapunov exponent: review on applications. Journal of Physics A: Mathematical and Theoretical, 46(25), 254019, 2013.

\bibitem[Chekroun et al.(2015a)]{Chekroun2015a}
Chekroun, M.D., Liu, H., Wang, S.: Approximation of Stochastic
Invariant Manifolds, , Springer, Cham, 2015a.

\bibitem[Chekroun et al.(2015b)]{Chekroun2015b}
Chekroun, M.D., Liu, H., Wang S.: Stochastic Parametrizing Manifolds
and Non-Markovian Reduced Equations, Springer, Cham, 2015b.

\bibitem[Daido(1984)]{Daido}
Daido, H.: Coupling sensitivity of chaos: a new universal property of chaotic
dynamical systems. Progr. Theoret. Phys. Suppl.  79,  p. 75, 1984.

\bibitem[De Cruz et al.(2016)]{Cruz2016}
De Cruz, L., Demaeyer, J., and Vannitsem, S.: The Modular Arbitrary-Order Ocean-Atmosphere Model: MAOOAM v1.0, Geosci. Model Dev., 9, 2793-2808, https://doi.org/10.5194/gmd-9-2793-2016, 2016.

\bibitem[De Cruz et al.(2018)]{DeCruz(2018)} De Cruz, L., Schubert, S., Demaeyer, J., Lucarini, V., and Vannitsem, S.: Exploring the Lyapunov instability properties of high-dimensional atmospheric and climate models, Nonlin. Processes Geophys., 25, 387-412, https://doi.org/10.5194/npg-25-387-2018, 2018. 

\bibitem[Eckmann and Ruelle(1985)]{Eckmann85}
Eckmann J.-P. and  Ruelle D.: Ergodic theory of chaos and strange attractors, Rev. Mod. Phys. {\bf 57}, 617, 1985.

\bibitem[Franzke et al.(2015)]{Franzke2015}
Franzke, C., Berner J., Lucarini V., OKane, T. J., Williams P. D.: Stochastic climate theory and modeling. Wiley Interdiscip. Rev.: Climate Change, 6, 6378, doi:https://doi.org/10.1002/wcc.318, 2015.

\bibitem[Fraedrich et al.(2005)]{Fraedrich2005}
Fraedrich, K., Kirk, E., Luksch, U., Lunkeit, F.: The portable university model of the atmosphere (PUMA): Storm track dynamics and low-frequency variability Meteorologische Zeitschrift Vol. 14 No. 6, p. 735 - 745, 2005.

\bibitem[Gallavotti and Lucarini(2014)]{Gallavotti2014} Gallavotti, G. and Lucarini, V.: Equivalence of nonequilibrium ensembles and representation of friction in  turbulent flows: the Lorenz 96 model. Journal of Statistical  Physics, 156 1027, 2014.

\bibitem[Ginelli et al.(2007)]{Ginelli07}
Ginelli, F. et al.: Characterizing Dynamics with Covariant Lyapunov Vectors, Phys. Rev. Lett. 99 130601, 2007.

\bibitem[Ginelli et al.(2013)]{Ginelli13}
Ginelli F. , Chat\'e H.,  Livi R.,  Politi A.: Covariant lyapunov vectors, J. Phys. A: Math. Theor. 46 254005, 2013.

\bibitem[Grassberger(1989)]{Grassberger89}
Grassberger P.: Information content and predictability of lumped and distributed dynamical systems, Phys. Scr. 40, 346, 1989.

\bibitem[Hallerberg et al.(2010)]{Hallerberg2010} Hallerberg S., Lopez, J. M., Pazo, D. and Rodriguez, M. A.: Logarithmic bred vectors in spatiotemporal chaos: Structure and growth. Phys. Rev. E, 81:066204, 2010.

\bibitem[Herrera et al.(2011)]{Herrera2011} Herrera S., Fern\'andez, J., Paz\'o, D., and Rodr\'iguez M. A.: The role of large-scale spatial patterns in the chaotic amplification of perturbations in a Lorenz’96 model, Tellus 63A, 978–990, 2011.

\bibitem[Karimi and Paul(2010)]{Karimi10}  Karimi A., Paul M. R.: Extensive chaos in the Lorenz-96 model, Chaos 20, 043105, 2010.

\bibitem[Kuptsov and Parlitz(2012)]{Kuptsov12}
Kuptsov P. V. and Parlitz U, J.: Theory and computation of covariant Lyapunov vectors, Nonlinear Sci. 22 727, 2012.

\bibitem[Livi et al.(1986)]{Livi86}
Livi R.,  Politi A., and Ruffo S.: Distribution of characteristic exponents in the thermodynamic limit, J. Phys. A 19, 2033, 1986.

\bibitem[Lorenz(1996)]{Lorenz96}
Lorenz, E. N.:Predictability: A problem partly solved. In ECMWF Seminar Proceedings I, Vol 1, 1995.

\bibitem[Lucarini and Sarno(2011)]{Lucarini2011} Lucarini V.,  Sarno S., Nonlin. Processes Geophys.: A statistical mechanical approach for the computation of the climatic response to general forcings, 18, 7–28, 2011.

\bibitem[Lucarini et al.(2014)]{L2014} Lucarini, V., Blender R., Herbert C., Pascale S., Ragone F., and Wouters J.: Mathematical and physical ideas for climate science, Rev. Geophys., 52, 809859, 2014.

\bibitem[Molteni(2003)]{Molteni2003}
Molteni, F.: Atmospheric simulations using a GCM with simplified physical  parametrizations. I. Model climatology and variability in multi-decadal experiments. Clim Dyn 20: 175-191, 2003. 

\bibitem[Mori et al.(1974)]{Mori1974}
Mori, H., Fujisaka, H., Shigematsu, H.: A new expansion of the master equation. Prog. Theor. Phys. 51: 109122, 1974.

\bibitem[Norwood et al.(2013)]{Norwood2013} 
Norwood A., Kalnay E., Ide K., Yang S.-C., Wolfe C., J. Phys. A, {\bf 46} 254021 (2013).

\bibitem[Orrel(2003)]{Orrell2003} Orrell, D.: Model Error and predictability over Different Timescales in the Lorenz96 System. J. Atmos. Sci. 60, 2219-2228, 2003.

\bibitem[Palmer and Williams(2008)]{Palmer2008}
Palmer, T.N., Williams, P.D.: Introduction. Stochastic physics and
climate modelling. Philos. Trans. Ser. A, Math., Phys., Eng. Sci. 366:
24212427, 2008.

\bibitem[Pavlioti et al.(2008)]{Pavliotis08}
Pavliotis, G. A., Stuart A. M., Multiscale Methods: Averaging and Homogenization, Springer, New York, 2008.

\bibitem[Peixoto and Oort(1992)]{Peixoto1992}
Peixoto, J., Oort, A.: Physics of Climate (American Institute of Physics, New
York), 1992.

\bibitem[Pikovsky and Politi(2016)]{PikovskyPoliti_LE} Pikovsky, A., Politi, A.: Lyapunov exponents: a tool to explore complex dynamics. Cambridge University Press, 2016.

\bibitem[Pugh et al.(2004)]{Pugh04}
Pugh, C., Shub, M., and Starkov, A.: Stable ergodicity, Bull. Am. Math. Soc. 41, 1, 2004.

\bibitem[Ruelle(1979)]{Ruelle79}
Ruelle, D.: Ergodic theory of differentiable dynamical systems, Publ. Math. IHES 50, 27, 1979.

\bibitem[Ruelle(1978)]{Ruelle78}
Ruelle, D.: Thermodynamic Formalism (Addison and Wesley, Reading, MA), 1978.

\bibitem[Shimada and Nagashima(1979)]{Shimada79}
Shimada, I. and Nagashima, T.: A numerical approach to ergodic problem of dissipative dynamical systems, Prog. Theor. Phys. 61 1605, 1979.

\bibitem[Takeushi et al.(2011)]{Kaz11}
Takeuchi, K. A. , Chat\'e,  H., Ginelli, F., Radons, G., Yang, H.-l.: Hyperbolic decoupling of tangent space and effective dimension of dissipative systems, Phys. Rev. E 84, 046214, 2011.

\bibitem[Trevisan and Uboldi(2004)]{Trevisan2004} Trevisan, A., and Uboldi, F.: Assimilation of standard and targeted observations
within the unstable subspace of the observation-analysis-forecast cycle. J. Atmos.
Sci. 61, 103-113, 2004.

\bibitem[Trevisan et al.(2010)]{Trevisan2010} Trevisan, A., D’Isidoro, M., Talagrand, O.: Four-dimensional variational assimilation
in the unstable subspace and the optimal subspace dimension. Quart. J. Roy.
Meteor. Soc. 136, 487-496, 2010.

\bibitem[Vannitsem and Lucarini (2016)]{Vannitsem2016}
Vannitsem, S. and Lucarini, V.: Statistical and dynamical properties of covariant lyapunov vectors in a coupled atmosphere-ocean model—multiscale effects, geometric degeneracy, and error dynamics, J. Phys. A: Math. Theor. 49 224001, 2016.

\bibitem[Vissio and Lucarini(2017)]{VissioLucarini}
Vissio G., Lucarini V.: A proof of concept for scale adaptive parametrizations: the case of the Lorenz '96 model, Quart. J. Roy. Meteor. Soc. 144 63, 2017.

\bibitem[Wilks(2006)]{Wilks2006} Wilks, D.S.: Effects of stochastic parametrizations in the Lorenz ’96 system,
Quart. J. Roy. Meteor. Soc. 131, 389-407, 2006.

\bibitem[Wouters and Lucarini(2012)]{Wouters2012}
Wouters J., Lucarini V.: Disentangling multi-level systems: Averaging,
correlations and memory. J. Stat. Mech: Theory Exp. 2012: P03003, 2012.

\bibitem[Wouters and Lucarini(2013)]{Wouters2013}
Wouters J., Lucarini V.: Multi-level dynamical systems: Connecting the
Ruelle response theory and the MoriZwanzig approach. J. Stat. Phys. 151:
850860, 2013.

\bibitem[Wouters et al.(2017)]{Wouters2017}
Wouters J.,  Gottwald G. A.: Edgeworth expansions for slow-fast systems with finite time scale separation, arXiv:1708.06984, 2017.

\bibitem[Yang et al.(2009)]{Kaz09}
Yang H.-l., Ginelli F., Chat\'e  H., Radons G., and Takeuchi K. A.: Hyperbolicity and the effective dimension of spatially extended dissipative systems,
Phys. Rev. Lett. 102, 074102, 2009.

\bibitem[Zwanzig(1960)]{Zwanzig1960}
Zwanzig R.: Ensemble method in the theory of irreversibility. J. Chem.
Phys. 33: 13381341, 1960.

\bibitem[Zwanzig(1961)]{Zwanzig1961}
Zwanzig R.: Memory effects in irreversible thermodynamics. Phys. Rev.
124: 983, 1961.






\end{thebibliography}

\end{document}